\documentclass[prd, reprint, amsfonts, amssymb, amsmath, preprintnumbers, showpacs, nofootinbib, superscriptaddress]{revtex4-1}
\usepackage{graphicx, amsthm, multirow}
\usepackage[citecolor=blue]{hyperref}
\usepackage[all]{hypcap}
\usepackage{url}
\usepackage{color}
\usepackage{subfigure}
\newcommand{\equaref}[1]{Eq.~(\ref{#1})}
\newcommand{\equasref}[2]{Eqs.~(\ref{#1})~and~(\ref{#2})}
\newcommand{\figref}[1]{Fig.~\ref{#1}}

\newcommand{\secref}[1]{Section~\ref{#1}}

\newcommand{\tabref}[1]{Table~\ref{#1}}
\newcommand{\refref}[1]{Ref.~\cite{#1}}

\begin{document}

\title{Testing atmospheric mixing sum rules at precision neutrino facilities}

\author{Peter Ballett}
\email{peter.ballett@durham.ac.uk}
\affiliation{IPPP, Department of Physics, Durham University, South Road, Durham DH1 3LE, United Kingdom}

\author{Stephen F. King}
\email{s.f.king@soton.ac.uk}
\affiliation{School of Physics and Astronomy, University of Southampton, Highfield, Southampton SO17 1BJ, United Kingdom}

\author{Christoph Luhn}
\email{christoph.luhn@durham.ac.uk}
\affiliation{IPPP, Department of Physics, Durham University, South Road, Durham DH1 3LE, United Kingdom}

\author{Silvia Pascoli}
\email{silvia.pascoli@durham.ac.uk}
\affiliation{IPPP, Department of Physics, Durham University, South Road, Durham DH1 3LE, United Kingdom}

\author{Michael A.~Schmidt}
\email{michael.schmidt@unimelb.edu.au}
\affiliation{ARC Centre of Excellence for Particle Physics at the Terascale, School of Physics, The University of Melbourne, Victoria 3010, Australia}

\date{\today}

\begin{abstract} 

We study the prospects for testing classes of \emph{atmospheric mixing sum
rules} at precision neutrino facilities. Such sum rules, which correlate the
atmospheric mixing angle $\theta_{23}$ with the recently measured reactor angle
$\theta_{13}$ and the cosine of the oscillation phase $\delta$, are predicted
by a variety of semi-direct models based on discrete family symmetry,
classified in terms of finite von Dyck groups.  We perform a detailed
simulation of the performance of the next generation of oscillation
experiments, including the wide band superbeam and low-energy neutrino
factory proposals, and compare their discriminating power for testing
\emph{atmospheric mixing sum rules}.

\end{abstract} \preprint{IPPP/13/64, DCPT/13/128} 

\maketitle

%%%%%%%%%%%%%%%%%%%%%%%%%%%%%%%%%%%%%%%%%%%%%%%%%%%%%%%%%%%%%%%%%%%%%%%%%%%%%%%
%%%%%%%%%%%%%%%%%%%%%%%%%%%%%%%%%%%%%%%%%%%%%%%%%%%%%%%%%%%%%%%%%%%%%%%%%%%%%%%
%%%%%%%%%%%%%%%%%%%%%%%%%%%%%%%%%%%%%%%%%%%%%%%%%%%%%%%%%%%%%%%%%%%%%%%%%%%%%%%
%%%%%%%%%%%%%%%%%%%%%%%%%%%%%%%%%%%%%%%%%%%%%%%%%%%%%%%%%%%%%%%%%%%%%%%%%%%%%%%
%%%%%%%%%%%%%%%%%%%%%%%%%%%%%%%%%%%%%%%%%%%%%%%%%%%%%%%%%%%%%%%%%%%%%%%%%%%%%%%
%%%%%%%%%%%%%%%%%%%%%%%%%%%%%%%%%%%%%%%%%%%%%%%%%%%%%%%%%%%%%%%%%%%%%%%%%%%%%%%

\section{Introduction} 
The recent measurement of the reactor mixing angle~$\theta_{13}$, by the Daya
Bay~\cite{An:2012eh,*An:2012bu} and RENO~\cite{Ahn:2012nd} experiments,
completes the measurement of the mixing angles in the
Pontecorvo-Maki-Nakagawa-Sakata (PMNS) matrix after the first hints
which appeared in 2011~\cite{Abe:2011sj,*Adamson:2011qu,*Abe:2011fz}.
The reactor angle $\theta_{13}$ turns out to be sizable,
$\sin^2(2\theta_{13})=0.089\pm0.010\pm0.005$~\cite{An:2012bu}, close to the
upper bound of the CHOOZ experiment~\cite{Apollonio:2002gd}. This discovery has
ruled out many of the most popular models of lepton flavour, which predicted
small or even vanishing $\theta_{13}$ at leading order.
Attention is now focused on models which can naturally incorporate the
large value of $\theta_{13}$. However, many such models do not predict this
angle uniquely, but instead predict \emph{atmospheric mixing sum rules}, where
the deviation of the atmospheric angle from its maximal value is controlled by
the product of the sine of the reactor angle and the cosine of the
oscillation phase $\delta$. 
The testability of these atmospheric mixing sum rules at future precision
neutrino facilities forms the subject of the present paper. 

At the time of writing, five parameters describing the neutrino sector
have been measured: three mixing angles and two mass-squared differences.  The
magnitude of CP violating effects in the lepton sector remains unknown, along
with the sign of the largest mass-squared difference. 
One final degree of freedom is given by the absolute neutrino mass scale, which
is bounded from above in the $1$~eV region by the results of tritium beta decay
experiments as well as cosmological and astrophysical
data~\cite{Beringer:1900zz}.
The mixing angles and phases constitute the PMNS matrix, which describes the
misalignment between flavour and mass bases. In the conventional
parameterization, it is expressed by 
\begin{align*} U_\text{PMNS} =~& \left ( \begin{matrix} 1 & 0 & 0 \\ 0 &
c_{23} & s_{23} \\ 0 & -s_{23} & c_{23} \end{matrix}\right)\left(
\begin{matrix} c_{13} & 0 & s_{13}e^{-\mathrm{i}\delta}\\ 0 & 1 & 0 \\
-s_{13}e^{\mathrm{i}\delta} & 0 & c_{13} \end{matrix}\right ) \\ &\times \left
( \begin{matrix} c_{12} & s_{12} & 0 \\ -s_{12} & c_{12} & 0 \\ 0 & 0 & 1
\end{matrix}\right) \left(\begin{matrix} 1 & 0 & 0 \\ 0 &
e^{\mathrm{i}\frac{\alpha_{21}}{2}} & 0 \\ 0 & 0 &
e^{\mathrm{i}\frac{\alpha_{31}}{2}} \end{matrix}\right)  \end{align*}
where $s_{ij}=\sin\theta_{ij}$, $c_{ij}=\cos\theta_{ij}$ and $\alpha_{ij}$ are
the two possible Majorana phases.  
The current $3\sigma$ intervals for the parameters of the neutrino sector
have been determined in a recent global analysis of oscillation
data~\cite{GonzalezGarcia:2012sz} to be
\begin{align*} \theta_{12} = [31^\circ&, 36^\circ],\quad \theta_{13} =
[7.2^\circ, 10^\circ], \quad \theta_{23} = [36^\circ, 55^\circ], \\ \Delta
m^2_{21} &= [7.00, 8.09]\times 10^{-5}~\text{eV}^2, \\ \Delta m^2_{31} &=
[2.27, 2.70]\times10^{-3}~\text{eV}^2~~~~~~~\,\text{(NO)}, \\ \Delta m^2_{32}
&= [-2.65, -2.24]\times10^{-3}~\text{eV}^2~~~\text{(IO)}, \end{align*} 
with NO and IO denoting normal and inverted neutrino mass
ordering.\footnote{For similar but independent global fits to neutrino
oscillation data, see~\cite{Tortola:2012te,*Fogli:2012ua}.} The phase $\delta$,
which enters the oscillation formulae through subdominant terms, is currently
unconstrained at $3\sigma$. The Majorana phases are also unconstrained but, as
they do not enter the neutrino oscillation formulae, must be addressed by
alternative experiments.

The large atmospheric and solar mixing angles evident in the leptonic sector
have motivated a number of authors to consider the existence of an underlying
(discrete) symmetry which connects states of different flavour. Approaches of
this type typically generate first-order expressions for the PMNS matrix which
are populated by simple algebraic values, and a number of such patterns have
been proposed, see~\refref{Altarelli:2010gt,*Ishimori:2010au,*King:2013eh} for
reviews with extensive lists of references. 
The simplest mixing patterns of this kind involve maximal atmospheric mixing
and a zero reactor angle, differing only in the solar mixing angle $\theta_{12}$. 
For example the tri-bimaximal (TB) mixing matrix \cite{Harrison:2002er}: 
\begin{equation}U_{\text{TB}} = \left(\begin{matrix} \sqrt{\frac{2}{3}} &
\frac{1}{\sqrt{3}} & 0\\ -\frac{1}{\sqrt{6}} & \frac{1}{\sqrt{3}} &
\frac{1}{\sqrt{2}} \\ \frac{1}{\sqrt{6}} & -\frac{1}{\sqrt{3}} &
\frac{1}{\sqrt{2}} \end{matrix}\right)P,\label{PMNS-TB}\end{equation}
predicted $\sin \theta_{12}=1/\sqrt{3}$. Another example, referred to as
golden ratio (GR) mixing, is given by the following
matrix~\cite{Datta:2003qg,*Kajiyama:2007gx}:    
\begin{equation} U_{\text{GR}} = \left(\begin{matrix}\cos\vartheta &
\sin\vartheta & 0 \\ -\frac{1}{\sqrt{2}}\sin\vartheta &
\frac{1}{\sqrt{2}}\cos\vartheta & \frac{1}{\sqrt{2}} \\
\frac{1}{\sqrt{2}}\sin\vartheta & -\frac{1}{\sqrt{2}}\cos\vartheta &
\frac{1}{\sqrt{2}} \\ \end{matrix}\right)P,\label{PMNS-GR}\end{equation}
where $\tan\vartheta = 1/\varphi$, with $\varphi$ given by the golden ratio
$(1+\sqrt{5})/2$. Although these are both excluded by the observation of the
reactor angle, the first or second columns of these matrices may be preserved
in the presence of a non-zero reactor angle, as we now discuss.

In the framework of so-called ``direct models'', both the mixing patterns
above have been shown to arise from some discrete family symmetry group G$_f$
(for example A$_4$, S$_4$ or
A$_5$)~\cite{Altarelli:2010gt,Ishimori:2010au,King:2013eh}. These are small
finite groups with three-dimensional representations, and frequently, 
the three generations of leptonic SU($2$) doublets are assigned to a triplet
representation, ensuring that their mixing is highly constrained. New scalar
fields are then introduced, called \emph{flavons}, which are also assigned to
representations of G$_f$, but are typically neutral under the standard model gauge
group. The lagrangian can then be written down in the conventional fashion,
with all terms included that are consistent with the symmetries of the
theory. The terms which constitute the flavon-flavon interactions are referred
to as the flavon potential; in successful models the minimum of this potential
will require non-zero vacuum expectation values (VEVs) for a subset of the
flavon fields, a feature which will spontaneously break G$_f$. The PMNS mixing
matrix then results from the presence of residual symmetries.

The residual symmetry in the charged lepton sector is based on the generator $T$,
while that in the (Majorana) neutrino sector is called the Klein symmetry
based on the $Z_2$ generators $S$ and $U$, where all three generators are contained
inside G$_f$ in the ``direct'' models~\cite{King:2009ap,King:2013eh}. 
In order to switch on the reactor angle, a popular approach is to break only the $U$
generator, leading to the so-called ``semi-direct'' approach~\cite{King:2013eh}
where the surviving $S$ generator maintains a particular column of the 
original mixing matrix. This keeps the solar angle close to its desired value,
while allowing a non-zero reactor angle which is correlated with the 
deviations of the atmospheric angle from its maximal value, depending on the 
cosine of the oscillation phase. 

In this work, we focus on the experimental prospects of constraining
generalized versions of such correlations known as \emph{atmospheric mixing
sum rules}: relations between the atmospheric mixing angle $\theta_{23}$ and
the recently measured reactor angle $\theta_{13}$. 
We shall show that these can describe a wide range of semi-direct models in
the literature and, with the increased sensitivity of the next generation of
oscillation experiments, will be significantly constrained for the first time
over the next few decades. After a study of the compatibility of different sum
rules with the current experimental results, as well as the projected
sensitivity of the extant experimental programme, we study two different
experimental proposals explicitly, namely a wide-band superbeam (WBB) with a
long baseline of around $2300$~km as well as the low-energy neutrino factory
(LENF). 

Although focusing on atmospheric mixing sum rules, this work will also
be relevant to the study of other types of correlations which are associated
with models of flavour. For example, solar mixing sum rules, which connect
$\theta_{12}$ to $\theta_{13}$ and
$\cos\delta$~\cite{King:2005bj,*Masina:2005hf,*Antusch:2005kw,*Antusch:2007rk},
can be associated with models of discrete symmetries where the leading-order
mixing pattern receives corrections from the charged lepton sector. The ability
to constrain these sum rules relies on the attainable precision on
$\theta_{12}$, and this will be set by the future medium-baseline reactor
experiments, JUNO and RENO-50, which predict an accuracy of below
$1$\%~\cite{Seo:2013aa, *Wang:2013aa}. However, to constrain $\cos\delta$, the
precision of the long-baseline physics programme, as considered in this work,
will be essential. 

%%%%%%%%%%%%%%%%%%%%%%%%%%%%%%%%%%%%%%%%%%%%%%%%%
%%%%%%%%%%%%%%%%%%%%%%%%%%%%%%%%%%%%%%%%%%%%%%%%%
%%%%%%%%%%%%%%%%%%%%%%%%%%%%%%%%%%%%%%%%%%%%%%%%%
%%%%%%%%%%%%%%%%%%%%%%%%%%%%%%%%%%%%%%%%%%%%%%%%%
%%%%%%%%%%%%%%%%%%%%%%%%%%%%%%%%%%%%%%%%%%%%%%%%%
%%%%%%%%%%%%%%%%%%%%%%%%%%%%%%%%%%%%%%%%%%%%%%%%%

The paper is organized as follows.  In~\secref{sec:sum rules}, we discuss the
sum rules arising in different classes of models. Technical group theoretical
details are deferred to the appendix. \secref{sec:higherOrders} addresses the
validity of the linearization approximation. The current experimental
constraints on the sum rules and the projected sensitivity of the current
experimental programme are discussed in~\secref{sec:predict}, while prospects
of next-generation experiments are presented in~\secref{sec:nextgen}. Finally,
we make our concluding remarks in~\secref{sec:conclusions}. 

%%%%%%%%%%%%%%%%%%%%%%%%%%%%%%%%%%%%%%%%%%%%%%%%%%%%%%%%%%%%%%%%%%%%%%%%%%%%%%%
%%%%%%%%%%%%%%%%%%%%%%%%%%%%%%%%%%%%%%%%%%%%%%%%%%%%%%%%%%%%%%%%%%%%%%%%%%%%%%%
%%%%%%%%%%%%%%%%%%%%%%%%%%%%%%%%%%%%%%%%%%%%%%%%%%%%%%%%%%%%%%%%%%%%%%%%%%%%%%%
%%%%%%%%%%%%%%%%%%%%%%%%%%%%%%%%%%%%%%%%%%%%%%%%%%%%%%%%%%%%%%%%%%%%%%%%%%%%%%%
%%%%%%%%%%%%%%%%%%%%%%%%%%%%%%%%%%%%%%%%%%%%%%%%%%%%%%%%%%%%%%%%%%%%%%%%%%%%%%%
%%%%%%%%%%%%%%%%%%%%%%%%%%%%%%%%%%%%%%%%%%%%%%%%%%%%%%%%%%%%%%%%%%%%%%%%%%%%%%%

\section{\label{sec:sum rules}Discrete family symmetries and sum rules}

In general, the incorporation of discrete family symmetries into any extension of the
standard model can only further our understanding of flavour if it manages to
reduce the number of free parameters in the theory. It is, therefore, generally
expected for these models to generate correlations amongst the physical
parameters governing the leptonic Yukawa sector. 
For a given model based on discrete family symmetries, the correlations
between the PMNS matrix elements will, in general, correspond to a
\emph{non-linear} relation amongst the mixing angles and phases. 

It is convenient to parameterize these relations by employing the notation of
\refref{King:2007pr}, which introduces the parameters $s$, $r$ and $a$
defined by 
\[ \sin\theta_{12} \equiv \frac{1+s}{\sqrt{3}}, \quad \sin\theta_{13} \equiv
\frac{r}{\sqrt{2}},\quad \sin\theta_{23}\equiv\frac{1+a}{\sqrt{2}}. \]
These parameters, which describe the deviations from tri-bimaximality, provide
a close phenomenological fit to the known mixing angles. The recent global fit
in~\refref{GonzalezGarcia:2012sz} provides the following $1\sigma$ intervals
(for normal neutrino mass ordering)
\begin{align*}  -0.07 \le s& \le -0.02,\\ 0.20 \le r& \le 0.23,\\ -0.12 \le a&
\le -0.05.  \end{align*}
In this paper, we will focus on a specific set of correlations which are
primarily dependent on the atmospheric mixing angle $\theta_{23}$, the reactor
mixing angle $\theta_{13}$ and the cosine of the Dirac CP phase, $\cos\delta$.
It will be useful to work with the first-order expansion of the exact
relation in the small parameters $s$, $r$ and $a$, which we call the
\emph{sum rule}. For the models that we are interested in, these will
take the general form
\begin{equation} a= a_0 + \lambda r \cos\delta + \mathcal{O}(r^2,
a^2),\label{general-sum rule} \end{equation}
where we will treat $a_0$ and $\lambda$ as new model-dependent constants. As
the mixing angles have already been measured, the sum rule can be
used to predict the Dirac CP phase $\delta$.  For phenomenologically viable
models, $a_0$ will always be small, of order of the $s$ parameter, and in the
analysis of \secref{sec:nextgen} it will be largely neglected. A
discussion of higher-order effects, correcting the sum rule, is
presented in \secref{sec:predict}.  

Although we will consider questions based on a range of
values of $\lambda$, there are two values which we would like to highlight.
These two choices have a degree of universality, having arisen in the
literature from fully consistent models, whilst also remaining the only simple
rules that we have found in our more phenomenological treatments: the first of
these rules has $\lambda=1$, and the second is given by $\lambda=-1/2$.
We will now illustrate this discussion with a few examples from the literature
which will highlight these two important cases.  A recent model presented in
\refref{King:2011zj} imposes an A$_4$ symmetry, broken spontaneously by a set
of flavons, which leads to the second column of the PMNS mixing matrix fixed at its
tri-bimaximal value,
\begin{align*} \left|U_{e2}\right| =  \left|U_{\mu2}\right| =  \left|U_{\tau2}\right| 
= \frac{1}{\sqrt{3}}.  \end{align*}
The corresponding exact relation can be linearized in terms of the $s$, $r$ and $a$
parameters \cite{King:2007pr}, 
\begin{equation} a = -\frac{r}{2}\cos\delta, \label{sumrulem0.5}\end{equation}
which is a specific realization of our general rule, \equaref{general-sum rule}
with $a_0=0$ and $\lambda=-1/2$. A different sum rule has been found in
\refref{Antusch:2011ic}, once again by spontaneously breaking the group A$_4$;
however, in this model the first column of the PMNS matrix is fixed at its
tri-bimaximal value. This imposes the relations, 
\begin{align*} \left | U_{e1} \right | &= \sqrt{\frac{2}{3}}
  \quad\text{and}\quad \left|U_{\mu1}\right| =\left|U_{\tau1}\right| =\frac{1}{\sqrt{6}}.  \end{align*}
Using these relations to compute the sum rule, one finds \cite{Antusch:2011ic},
\begin{equation} a = r\cos\delta, \label{sumrule1}\end{equation}
which corresponds to $a_0=0$ and $\lambda=1$.

%%%%%%%%%%%%%%%%%%%%%%%%%%%%%%%%%%%%%%%%%%%%%%%%%%%%%%%%%%%%%%%%%%%%%%%%%%%%%%%
%%%%%%%%%%%%%%%%%%%%%%%%%%%%%%%%%%%%%%%%%%%%%%%%%%%%%%%%%%%%%%%%%%%%%%%%%%%%%%%
%%%%%%%%%%%%%%%%%%%%%%%%%%%%%%%%%%%%%%%%%%%%%%%%%%%%%%%%%%%%%%%%%%%%%%%%%%%%%%%
%%%%%%%%%%%%%%%%%%%%%%%%%%%%%%%%%%%%%%%%%%%%%%%%%%%%%%%%%%%%%%%%%%%%%%%%%%%%%%%
%%%%%%%%%%%%%%%%%%%%%%%%%%%%%%%%%%%%%%%%%%%%%%%%%%%%%%%%%%%%%%%%%%%%%%%%%%%%%%%
%%%%%%%%%%%%%%%%%%%%%%%%%%%%%%%%%%%%%%%%%%%%%%%%%%%%%%%%%%%%%%%%%%%%%%%%%%%%%%%

A novel method was recently introduced by Hernandez and Smirnov
in \refref{Hernandez:2012ra} which produces flavour-symmetric correlations
amongst the PMNS mixing matrix elements, whilst making minimal assumptions
about the details of the model.  This approach was built around the assumption
that there exists a discrete flavour group which is broken spontaneously into
two subgroups. These subgroups act independently on the charged lepton and
neutrino sectors of the theory, and their misalignment leads to a non-trivial
PMNS matrix. If we assume, in this framework, that some of the known symmetries
of the leptonic mass terms are in fact residual symmetries arising from this
larger broken group, constraints can be placed on the PMNS matrix in a general
manner, regardless of the precise implementation of the symmetry breaking. For
the groups that we will focus on, the constraints which arise from this
construction fix one column of the PMNS matrix: 
\begin{align*} \left|U_{\alpha i}\right|^2 = \eta,\qquad\text{and}\qquad\left
|U_{\beta i}\right|^2 = \left |U_{\gamma i} \right |^2 = \frac{1-\eta}{2},
\end{align*}
where $\{\alpha,\beta,\gamma\}=\{e,\mu,\tau\}$ and the parameter $\eta$ is a
model dependent constant, which can be found in the appendix.
Fixing a column of the PMNS matrix introduces two independent constraints on
the mixing angles. For the cases that we are interested in, either the first or
second column is fixed, and we can express these constraints as an exact
prediction for $s$ as a function of $r^2$, and an atmospheric sum rule of the
general form as given in~\equaref{general-sum rule}.

One can show that, working within this framework, there is a finite number of
possible values for $\eta$, depending on the underlying group and the choice of
generators preserved after spontaneous symmetry breaking. Using the exact
expressions for $s$, we can make predictions for each choice of $\eta$ and
exclude any models that are incompatible with the current experimental data. At
the end of this process, we are left with a finite number of models with
phenomenologically viable predictions of $s$ and a definite atmospheric
sum rule.
It turns out that these are all closely related to the two special sum rules
that we have already identified in \equasref{sumrulem0.5}{sumrule1}. Generally they
predict sum rules with values of $\lambda$ numerically close to $1$ or $-1/2$.
A full listing of these rules is given in \tabref{all-sumrules-numbers}, and we refer
the reader to the appendix for the details of their derivation. We see that by
choosing different residual generators, we find $8$ distinct sum rules of the
type of \equaref{general-sum rule} which are compatible with the current
phenomenological data. 

A number of the scenarios that we have identified in
\tabref{all-sumrules-numbers} can be explained in terms of the TB and GR
matrices given in \equasref{PMNS-TB}{PMNS-GR}. The three scenarios based on an
A$_4$ symmetry all lead to a value of the second column of the PMNS
matrix fixed at its tri-bimaximal value; similarly, the S$_4$ scenario with the
generator choice $T_e$--$S_1$ fixes the prediction of the first
column to be tri-bimaximal. The scenario based on A$_5$ with unbroken
generators $T_e$--$S_1$ ($T_e$--$S_2$) fixes the first (second) column of the
PMNS matrix to the equivalent values of the GR mixing matrix. 

\begin{table}
\centering
\begin{tabular}{c c c c c c}
$G_f$ & $m$ & $T_\alpha$,$S_i$ & $s$ & $a_0$ & $\lambda$\\
\hline\hline
\multirow{3}{*}{A$_4$} & $3$ & $T_e$,$S_2$ & $0.012$ & $0$ & $-0.5$ \\
& $3$ & $T_\mu$,$S_2$ & $0.012$ & $0$ & $-0.5$ \\
& $3$ & $T_\tau$,$S_2$ & $0.012$ & $0$ & $-0.5$ \\\hline
\multirow{3}{*}{S$_4$} & $3$ & $T_e$,$S_1$ & $-0.024$ & $0$ & $1$ \\
& $4$ & $T_\mu$,$S_2$ &$-0.124$ & $-0.167$ & $-0.408$ \\
& $4$ & $T_\tau$,$S_2$&$-0.124$ & $0.167$ & $-0.408$ \\\hline
\multirow{4}{*}{A$_5$}
& $5$ & $T_e$,$S_1$ &$-0.118$ & $0$ & $1.144$ \\
& $5$ & $T_e$,$S_2$ &$-0.079$ & $0$ & $-0.437$ \\
& $5$ & $T_\mu$,$S_2$ &$0.054$ & $0.067$ & $-0.532$ \\
& $5$ & $T_\tau$,$S_2$ &$0.054$ & $-0.067$ & $-0.532$ \\
\end{tabular}

\caption{\label{all-sumrules-numbers}The phenomenologically viable sum
rules of the form $a= a_0 + \lambda r \cos\delta $ (where $a,r$ are the
atmospheric and reactor angle deviations from tri-bimaximal mixing and $\delta$
is the CP violating oscillation phase) arising in the Hernandez-Smirnov
framework for finite von Dyck groups. In this table, $m$ gives the order of the
generator which controls the charged lepton mass matrix, $T^m_\alpha=1$, while
$S_i$ is the generator of the von Dyck group that is identified with one of the
generators of the Klein symmetry of the neutrino mass matrix (with the other
Klein symmetry generator being unrelated to the von  Dyck group, as in
so-called semi-direct models). Analytical expressions for the solar angle
deviation from tri-bimaximal mixing $s$ and the constants $a_0$ and $\lambda$
are given in \tabref{tab:all-sumrules}. The numerical values are obtained for
the current best-fit value of $\sin^22\theta_{13}=0.089$~\cite{An:2012bu}.}
\end{table}

%%%%%%%%%%%%%%%%%%%%%%%%%%%%%%%%%%%%%%%%%%%%%%%%%%%%%%%%%%%%%%%%%%%%%%%%%%%%%%%
%%%%%%%%%%%%%%%%%%%%%%%%%%%%%%%%%%%%%%%%%%%%%%%%%%%%%%%%%%%%%%%%%%%%%%%%%%%%%%%
%%%%%%%%%%%%%%%%%%%%%%%%%%%%%%%%%%%%%%%%%%%%%%%%%%%%%%%%%%%%%%%%%%%%%%%%%%%%%%%
%%%%%%%%%%%%%%%%%%%%%%%%%%%%%%%%%%%%%%%%%%%%%%%%%%%%%%%%%%%%%%%%%%%%%%%%%%%%%%%
%%%%%%%%%%%%%%%%%%%%%%%%%%%%%%%%%%%%%%%%%%%%%%%%%%%%%%%%%%%%%%%%%%%%%%%%%%%%%%%
%%%%%%%%%%%%%%%%%%%%%%%%%%%%%%%%%%%%%%%%%%%%%%%%%%%%%%%%%%%%%%%%%%%%%%%%%%%%%%%

\section{\label{sec:higherOrders}Validity of linearization}

\begin{figure}[t] \centering
\includegraphics[width=0.75\linewidth, angle=270]{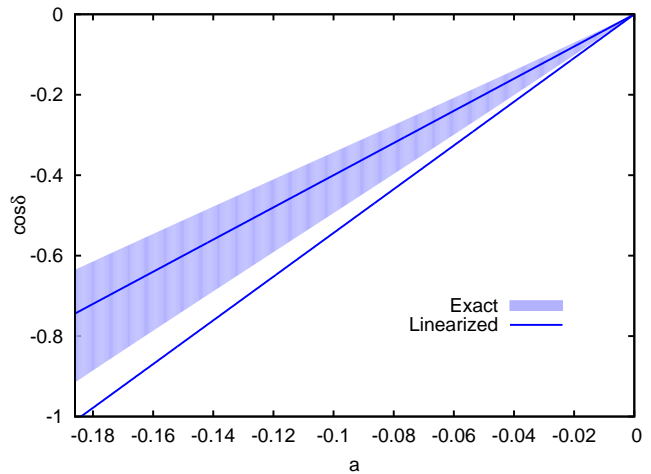} 
\caption{\label{lin-comp} A comparison between the exact correlation and the
sum rule for the model presented in \refref{Antusch:2011ic},
which fixes the elements of the first column of the PMNS matrix to their
tri-bimaximal values. The solid (empty) region denotes the exact (linearized)
prediction for $\cos\delta$ which is produced by varying $r$ over its current
$3\sigma$ allowed interval. } 
 \end{figure}

In general, the correlations predicted by flavour symmetric models are
non-linear relations between the oscillation parameters. 
We have discussed how the form of these correlations simplifies when
only the first-order terms in the parameters $s$, $r$ and $a$ are retained, and
we will now address the impact of higher-order terms.  We consider the model
presented in \refref{Antusch:2011ic}, which fixes the elements of the first
column of the PMNS matrix to their tri-bimaximal values. As a
function of $r$ and $a$, this model predicts that $\cos\delta$ is given by the
composition of the following functions:
\begin{align*} 
&\cos\delta = \frac{(-2\sin^2\theta_{12}+\cos^2\theta_{12}
r^2)\cos(2\theta_{23})}{\sqrt{2}r\sin(2\theta_{12})\sin(2\theta_{23})},\\
&~~\cos\theta_{12} = \frac{2}{\sqrt{3(2-r^2)}},\qquad \text{and}
\qquad\sin\theta_{23} = \frac{1+a}{\sqrt{2}}.  \end{align*}
When linearized, these relations lead to the simpler expression $\cos\delta =
a/r$.  In \figref{lin-comp} we have computed the predictions of $\cos\delta$ as
a function of $a$ for both the exact relation and the sum rule, with
$r$ varied within its experimentally allowed 3$\sigma$ region. We see that for
this model the difference between the two treatments is small. The impact of
higher order corrections can only be assessed on a case by case basis once the
exact correlations are known; however, due to the smallness of the
$r$ and $a$ parameters, we expect the linear approximation to be a good
one. This is confirmed by our simulations for the known exact
correlations, and therefore we will focus our later analysis on the linearized
relations. This also allows us to treat the universality that we have observed
in \secref{sec:sum rules}, all viable sum rules that we have identified are
either close to $\lambda=1$ or $\lambda=-1/2$. For the classes of
phenomenologically viable models that we have found, the differences
between similar sum rules are small and will be very
challenging to measure.\\

%%%%%%%%%%%%%%%%%%%%%%%%%%%%%%%%%%%%%%%%%%%%%%%%%%%%%%%%%%%%%%%%%%%%%%%%%%%%%%%
%%%%%%%%%%%%%%%%%%%%%%%%%%%%%%%%%%%%%%%%%%%%%%%%%%%%%%%%%%%%%%%%%%%%%%%%%%%%%%%
%%%%%%%%%%%%%%%%%%%%%%%%%%%%%%%%%%%%%%%%%%%%%%%%%%%%%%%%%%%%%%%%%%%%%%%%%%%%%%%
%%%%%%%%%%%%%%%%%%%%%%%%%%%%%%%%%%%%%%%%%%%%%%%%%%%%%%%%%%%%%%%%%%%%%%%%%%%%%%%
%%%%%%%%%%%%%%%%%%%%%%%%%%%%%%%%%%%%%%%%%%%%%%%%%%%%%%%%%%%%%%%%%%%%%%%%%%%%%%%
%%%%%%%%%%%%%%%%%%%%%%%%%%%%%%%%%%%%%%%%%%%%%%%%%%%%%%%%%%%%%%%%%%%%%%%%%%%%%%%

\section{\label{sec:predict}Compatibility of sum rules with existing and projected data}

\begin{figure*} \centering
\subfigure[~Normal mass ordering\label{fig:predictNH}]{\includegraphics[width=0.35\linewidth,angle=-90,origin=c,trim=0 110 0 50]{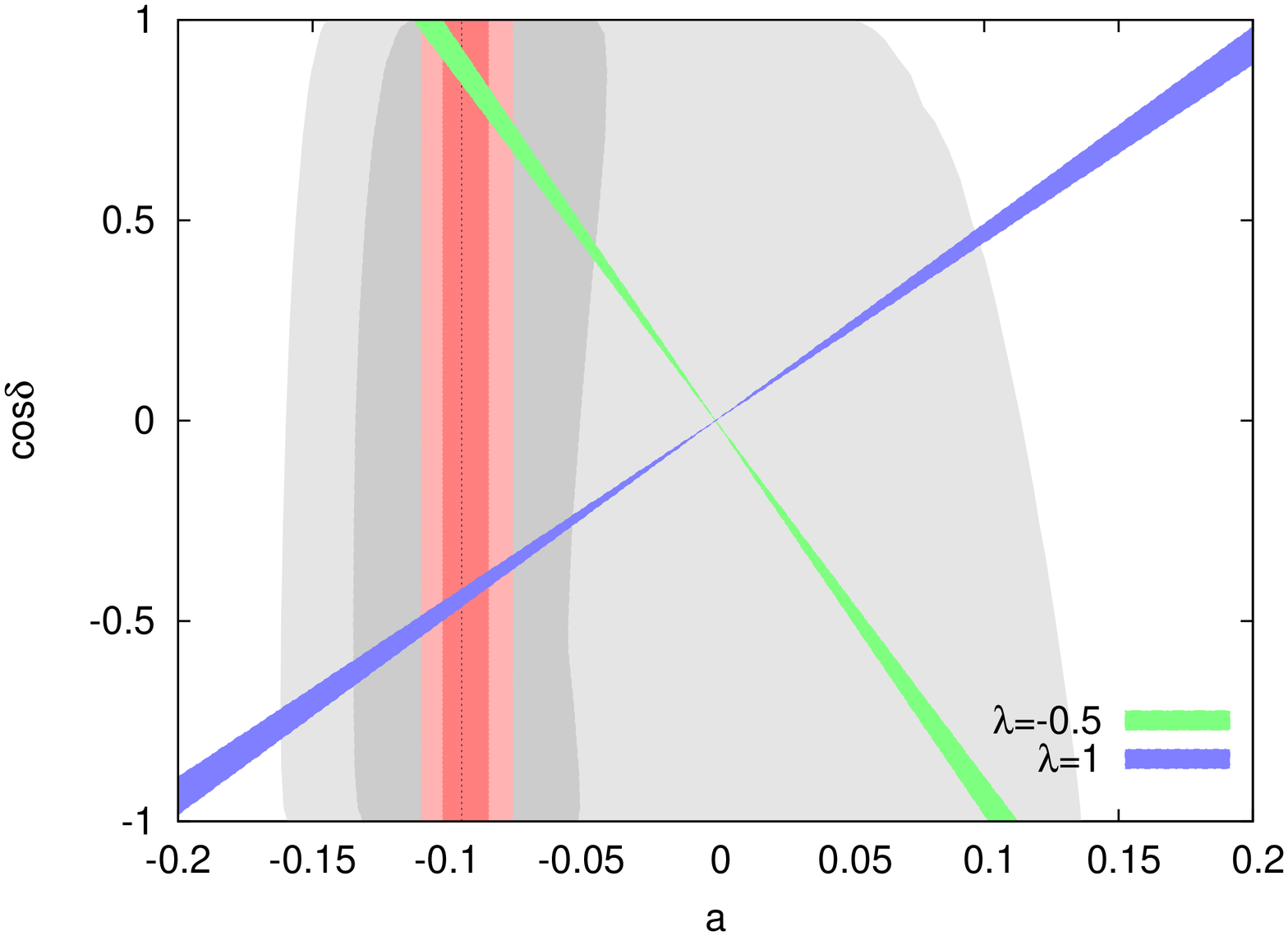} }
\hspace{1.5cm}
\subfigure[~Inverted mass ordering\label{fig:predictIH}]{\includegraphics[width=0.35\linewidth,angle=-90,origin=c,trim=0 110 0 50]{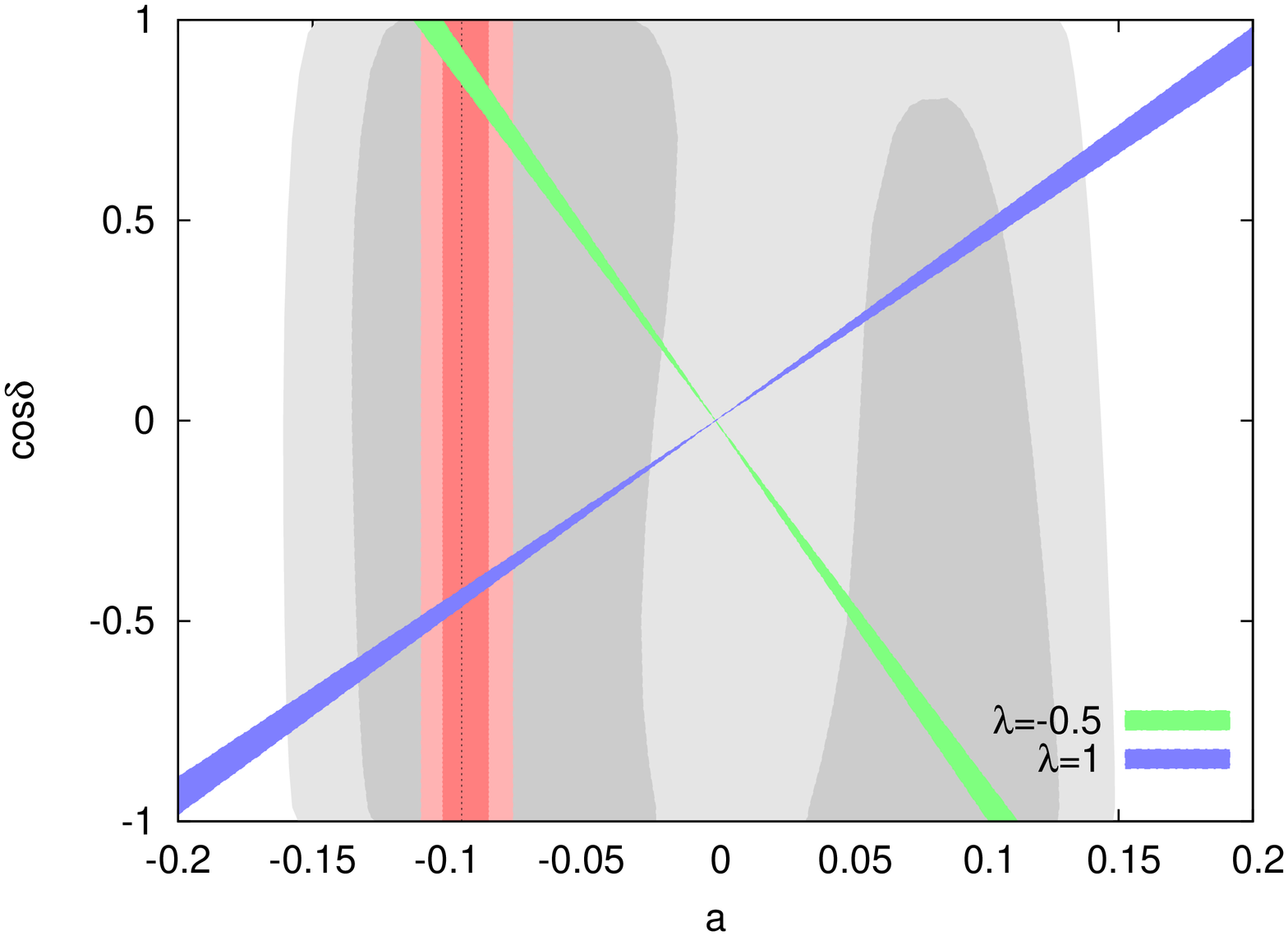} }
\caption{\label{predict} The current experimental status of the sum rules in
\equaref{general-sum rule} given by $\lambda=1$ and $\lambda=-0.5$, with
$a_0=0$. The diagonal lines show the regions predicted for $a$ and
$\cos\delta$ given the $3\sigma$ bounds on $r$, assuming both normal ordering
(\figref{fig:predictNH}) and inverted ordering (\figref{fig:predictIH}). The
vertical line shows the current best-fit for $a$ where the projected
sensitivity is indicated by the red bands; the dark (light) grey regions show
the current $1\sigma$ ($2\sigma$) allowed intervals~\cite{GonzalezGarcia:2012sz}.} 
\end{figure*}

The global neutrino oscillation data already constrains models which exhibit
discrete flavour symmetries. 
For a given model, our general sum rule can be used to predict the
value $\cos\delta$. Fixing $a$, we define $\cos\delta$ by the mapping from $r$
which is found by inverting \equaref{general-sum rule}; $r$ is then allowed to
vary across its $1\sigma$ interval \cite{GonzalezGarcia:2012sz} and the image
of this mapping is taken to be the range of potential values for $\cos\delta$.

In \figref{predict} we show the predictions of our two specific sum rules and
their compatibility with the current global data on $a$ (the grey regions). We
have also shown (the red bands) the projected sensitivity to the $a$ parameter
as reported in \refref{Huber:2009cw}. These projections are for the global
parameter sensitivity in 2025 assuming only the current experimental programme:
$5$ years of data from T2K, $6$ from NO$\nu$A, and $3$ years each for Double
Chooz, RENO and Daya Bay. As we cannot predict the future best-fit value, the
horizontal location of the predicted regions is largely irrelevant, and in
\figref{predict} they have been arbitrarily centred around the current best-fit
value.

We see that the predictions of $\delta$ for these two models are currently
consistent with the global data. However, the overlap for some of these $1\sigma$
intervals can be seen to require some quite specific correlations: for example,
$\lambda=-0.5$ and NO requires $\cos\delta\gtrsim0.5$. With the projected
sensitivity to $a$, these correlations could create tension with the future
data, and the consistency of these models will start to become rather
constrained. 
For example, in a strictly CP-conserving theory, $\sin\delta$ must vanish.
The corresponding value of $\cos\delta$ would then be difficult to reconcile with
the sum rule given by $\lambda=1$, leading to a possible exclusion of such a sum rule.
The limiting factor for the general exclusion of these models with the
current experimental programme will be the attainable precision on
$\cos\delta$. It has been shown that, in the most optimistic case, the current
experimental programme will only be able to provide a $3\sigma$ region for
$\delta$ with a width of around $300^\circ$\cite{Huber:2004gg}.   It is clear,
therefore, that testing mixing sum rules will be a task to be addressed by a
next-generation neutrino oscillation facility, one which focuses on precision.

%%%%%%%%%%%%%%%%%%%%%%%%%%%%%%%%%%%%%%%%%%%%%%%%%%%%%%%%%%%%%%%%%%%%%%%%%%%%%%%
%%%%%%%%%%%%%%%%%%%%%%%%%%%%%%%%%%%%%%%%%%%%%%%%%%%%%%%%%%%%%%%%%%%%%%%%%%%%%%%
%%%%%%%%%%%%%%%%%%%%%%%%%%%%%%%%%%%%%%%%%%%%%%%%%%%%%%%%%%%%%%%%%%%%%%%%%%%%%%%
%%%%%%%%%%%%%%%%%%%%%%%%%%%%%%%%%%%%%%%%%%%%%%%%%%%%%%%%%%%%%%%%%%%%%%%%%%%%%%%
%%%%%%%%%%%%%%%%%%%%%%%%%%%%%%%%%%%%%%%%%%%%%%%%%%%%%%%%%%%%%%%%%%%%%%%%%%%%%%%
%%%%%%%%%%%%%%%%%%%%%%%%%%%%%%%%%%%%%%%%%%%%%%%%%%%%%%%%%%%%%%%%%%%%%%%%%%%%%%%

\section{\label{sec:nextgen}Testing sum rules at next-generation facilities}

With the knowledge of the value of $\theta_{13}$ the campaign for a
next-generation facility, designed to make precision measurements of the
neutrino mixing parameters, is greatly strengthened. It is likely that within
the extant experimental neutrino physics programme, we will see hints towards the
measurement of two of the most important unknowns in the conventional neutrino
flavour-mixing paradigm: the sign of the atmospheric mass-squared
difference and the value of the CP-violating phase, $\delta$. It is, however,
unlikely that these questions will be resolved at an acceptable statistical
confidence level: the projected $3\sigma$ CP-violation discovery fraction with
the current experimental programme only reaches around $20\%$ of the parameter
space \cite{Huber:2009cw} and it is only modestly higher for the determination
of the mass ordering at around $40\%$. The desire for a definitive $5\sigma$
answer to these questions provides the first motivation for the construction
of a next-generation neutrino oscillation facility, capable of precision
measurements of the oscillation parameters.
In this work, we will focus on two such designs: the low-energy neutrino
factory (LENF) and a wide-band superbeam (WBB).

The WBB is an extrapolation of existing technology, using a more powerful
version of the conventional neutrino beam production method.  Protons are
accelerated towards a target, and the subsequent collision generates, amongst
other things, pions and kaons. Magnetic horns then focus the meson beam,
selecting $\pi^+$, the decay of which generates the neutrino beam,
predominantly composed of $\nu_\mu$ with a small contamination of $\nu_e$,
$\bar{\nu}_\mu$ and $\bar{\nu}_e$ which constitute a background to the signal,
with an analogous composition for the initial selection of $\pi^-$.
After years of experimental work on similar designs, this technology is very
well understood and considerable expertise is to be found in the community.
With a large value of $\theta_{13}$, such a next-generation superbeam has been
shown~\cite{Richter:2000pu, Coloma:2012ut} to provide a quite competitive
physics reach compared to other designs, and for a significant fraction of
parameter space, may be sensitive to CP violation originating from the PMNS
matrix.
There are a number of proposed experiments based on the WBB design. The CERN to
Pyh\"asalmi superbeam (C2P) has been developed, and recently recommended, by
the LAGUNA-LBNO design study
\cite{Autiero:2007zj,*Angus:2010sz,*Rubbia:2010fm,*Rubbia:2010zz}. For this
experiment neutrinos produced at CERN are detected by a $70$~kton liquid argon
detector (LAr) after a propagation distance of $2300$~km at
Phy\"asalmi in Finland.
There is a similar proposal for an intense long-baseline superbeam known as
LBNE, which is based in America. In its first phase, the facility consists of a
$10$~kton liquid argon detector based on the surface at Homestake, separated by
a distance of $1300$~km from Fermilab.  This should be viewed as the first step
in a staged programme, ultimately aiming for an underground detector of order
$35$~kton, which has been shown to have strong discovery potential for the mass
ordering and CP violation effects.

At a neutrino factory
\cite{Geer:1997iz,*DeRujula:1998hd,*Bandyopadhyay:2007kx}, a neutrino beam is
produced via the decay of muons held in a storage ring. This process is very
well understood and controlled, which leads to small systematic uncertainties
and ultimately strong sensitivity to the neutrino mixing parameters.
The typical design has evolved over the last few years. The original
designs worked with a high-energy facility, with stored-muon energies of around
$25$~GeV \cite{Choubey:2011zz}.  This was shown to have exceptional sensitivity
for small values of $\theta_{13}$, down to $\sin^22\theta_{13}=10^{-5}$
\cite{Huber:2006wb,*Agarwalla:2010hk}. However, with the discovery of the large
value of $\theta_{13}$ by the Daya Bay \cite{An:2012eh,*An:2012bu} and RENO
\cite{Ahn:2012nd} experiments, the consensus has now fallen on the low-energy
variant, the LENF \cite{Bross:2007ts}, designed with a stored-muon energy of
around $10$~GeV. At the LENF, a strong sensitivity to the PMNS parameters is
achieved by focusing on the rich oscillation signal which can be found in the
low-energy parts of the neutrino spectrum, a technique which relies on the
enhanced number of events associated with larger values of $\theta_{13}$.
Optimization work on the LENF has shown it to be a versatile design
\cite{FernandezMartinez:2010zza,Ballett:2012rz} and a strong candidate for a
precision neutrino oscillation facility \cite{Coloma:2012wq}.  Assuming $\mu^-$
($\mu^+$) in the storage ring, the beam of a neutrino factory consists of
$\nu_\mu~(\overline{\nu}_\mu)$ and $\overline{\nu}_e~(\nu_e)$. The LENF is
designed to focus primarily on the measurement of ``wrong-sign muons'', the
antiparticles of those in the storage ring, which are produced by
charged-current interactions in the detector after the flavour transition
$\overline{\nu}_e~(\nu_e) \to \overline{\nu}_\mu~(\nu_\mu)$.  In addition
to this ``golden channel'', it may also be possible to include the ``platinum
channel'', observing electrons at the detector produced by incident
$\overline{\nu}_e~(\nu_e)$ \cite{FernandezMartinez:2010zza}. This additional
channel is only available to certain detector technologies, and has been shown
to confer only a slight improvement for the traditional discovery searches;
however, its impact on precision measurements is as yet unknown. Nevertheless,
in this study we will not consider the impact of the platinum channel, only
assuming the observation of $\mu^+$ ($\mu^-$) at the detector.

A number of alternative detector options are often considered for the
LENF~\cite{FernandezMartinez:2010zza,Agarwalla:2010hk,Ballett:2012rz}. 
In this work, we are not looking to make a detailed comparison of designs, but
instead to show the feasibility of constraining sum rules at next-generation
facilities. As such, we have restricted our attention to two variants of the
LENF detector design: a magnetized iron neutrino detector (MIND) and a
magnetized liquid argon detector (mLAr) based on liquid argon time-projection
chamber technology. These two technologies provide us with a fair estimate of
performance of the current proposals (MIND), as well as a more optimistic
assessment (mLAr) of the potential of a LENF. 
A LENF with a MIND has become the favoured design of the International Design
Study for a Neutrino Factory~\cite{Long:2013aa}. 
The MIND is composed of alternating sheets of iron and scintillator, placed
inside in a $1.5$~T toroidal magnetic field. 
This technology is very well understood, based on an extrapolation of the MINOS
detector to a larger scale, and has been the object of extensive study,
demonstrating strong physics reach~\cite{Agarwalla:2010hk, Coloma:2012wq,
Coloma:2012ji, Blennow:2013swa}. 
A large magnetized liquid argon detector, would be an ultimate detector for a
LENF, as it allows for detailed event reconstruction, a low threshold energy
and excellent energy resolution. Such a facility, and in particular the
magnetization of the large detector volume, poses some technical challenges.
In our study, we simulate a scenario based on a $50$ kton mLAr which
should be viewed as providing an optimistic upper bound on the performance of
the LENF.  

\begin{figure*}[ht] \centering
\subfigure{\includegraphics[clip, trim=17 24 5 22, angle=270,width=.32\linewidth]{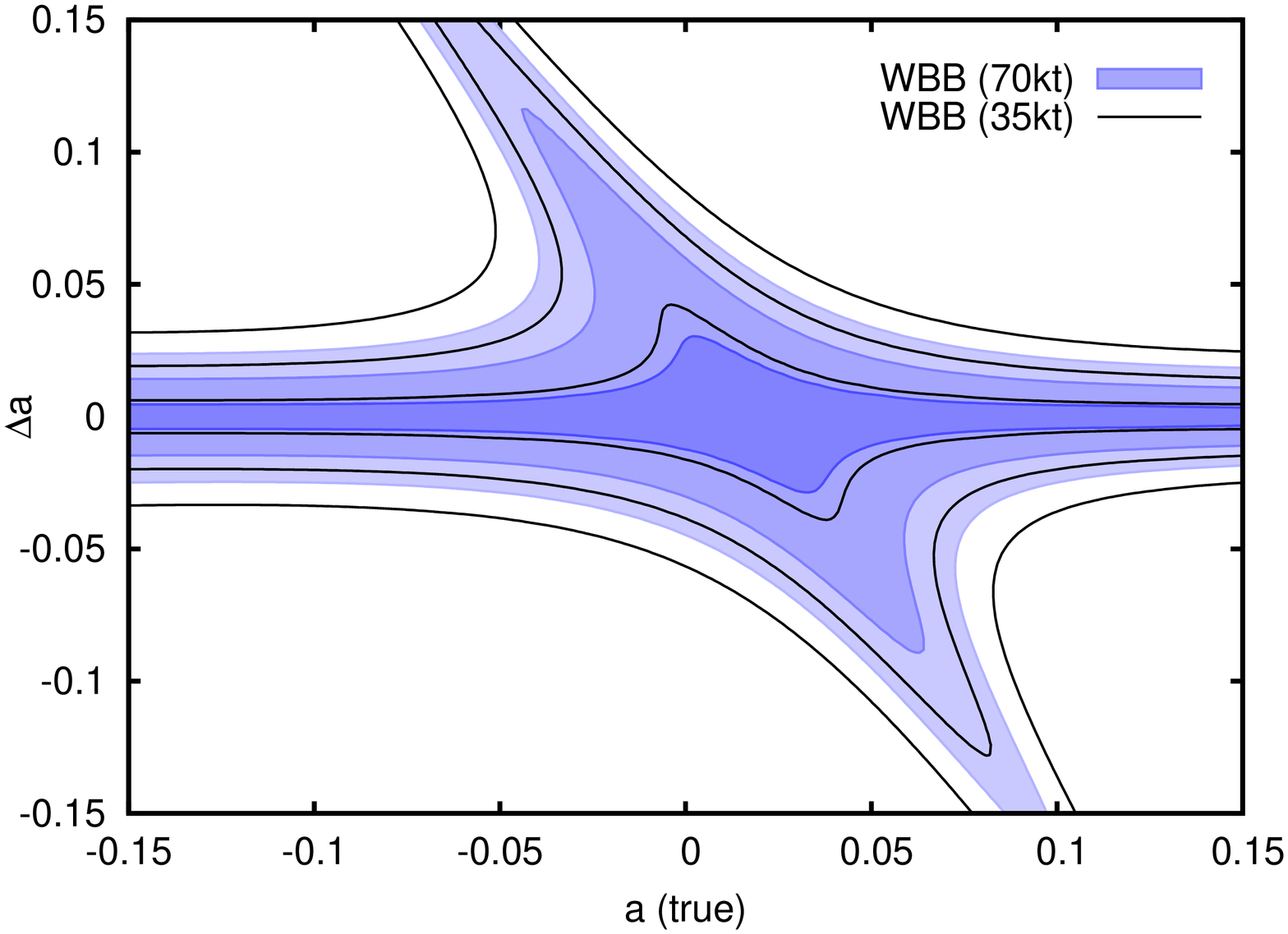}} 
\subfigure{\includegraphics[clip, trim=17 24 5 22, angle=270,width=.32\linewidth]{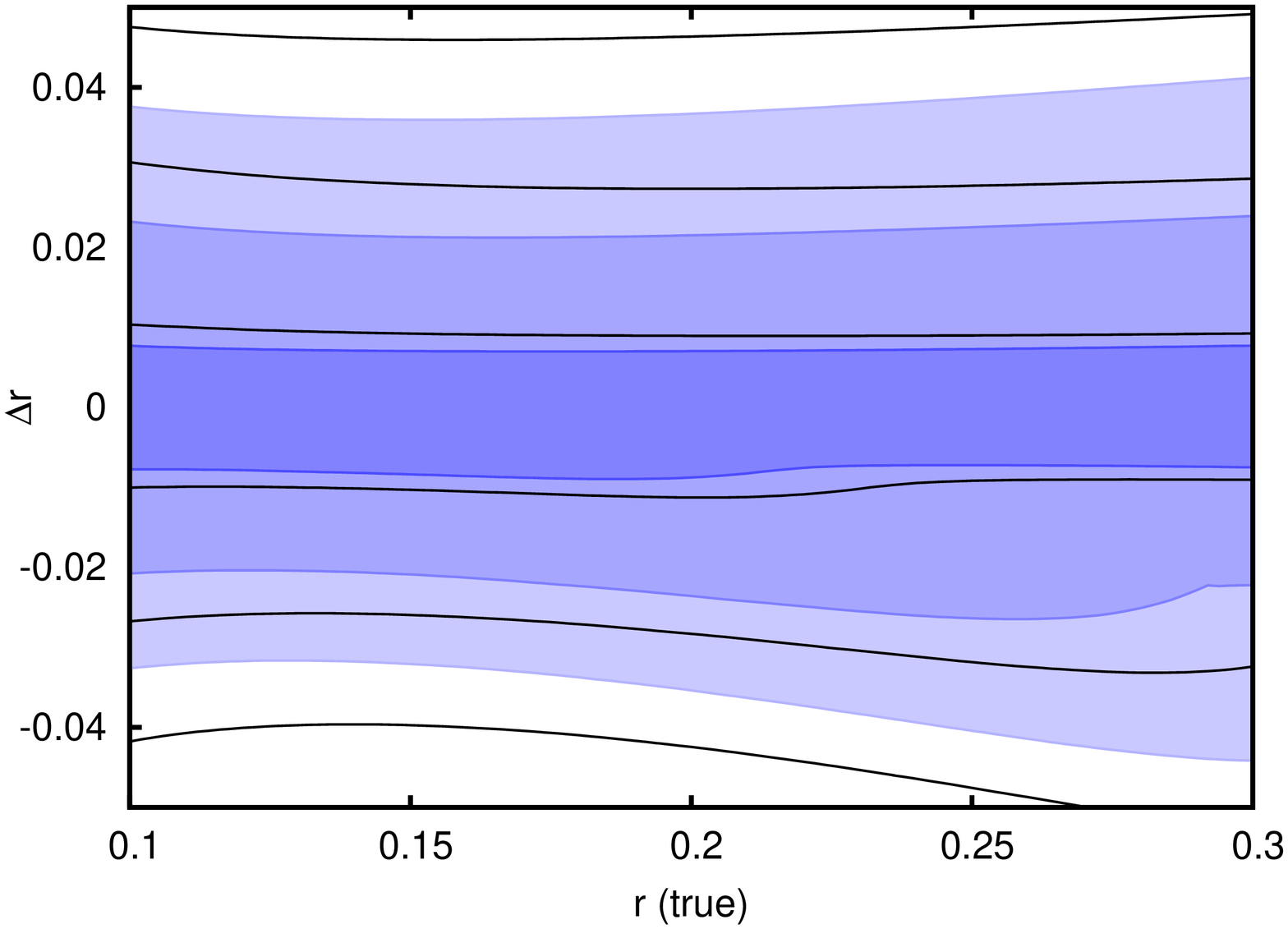}} 
\subfigure{\includegraphics[clip, trim=17 24 5 22, angle=270,width=.32\linewidth]{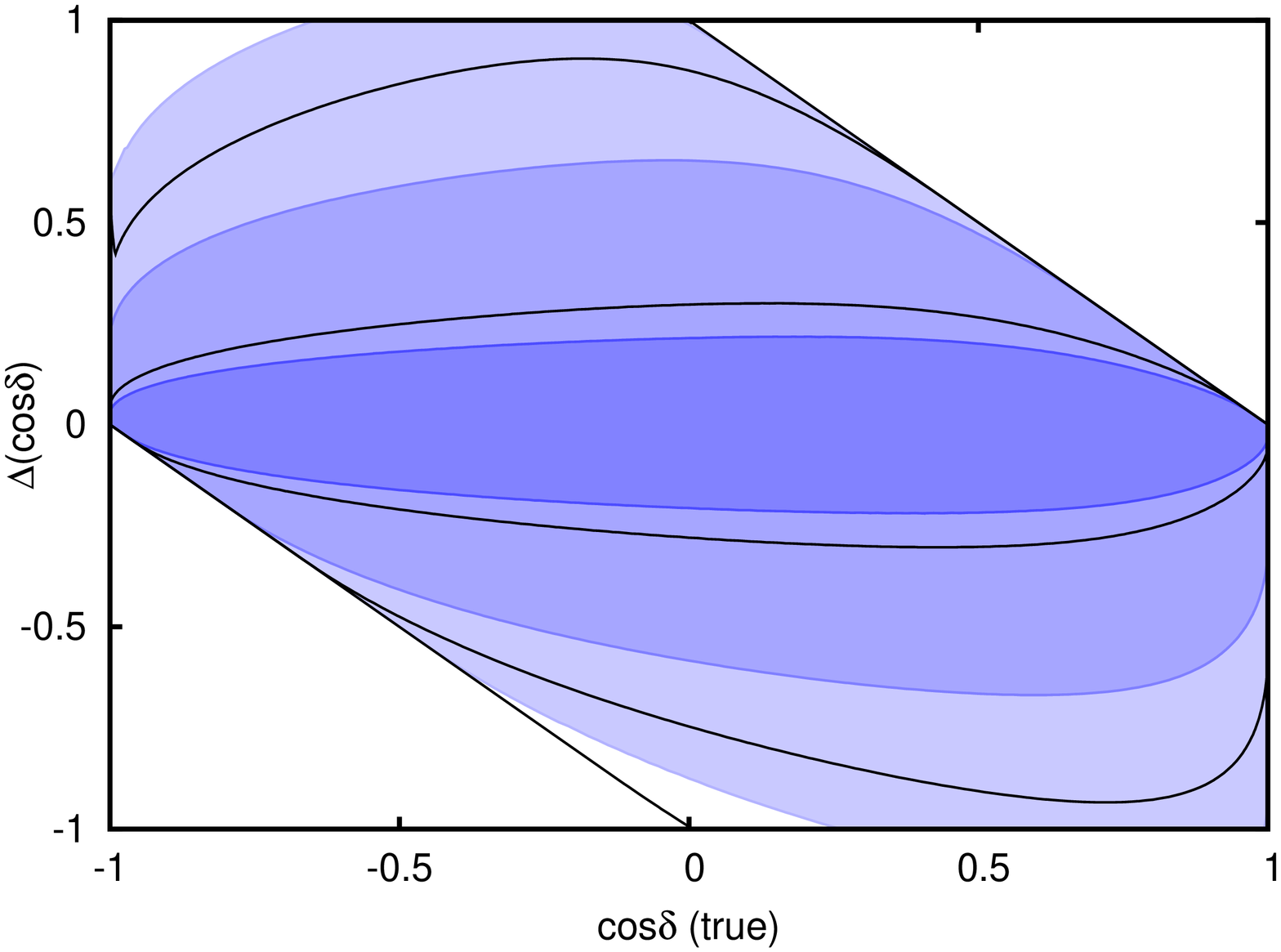}} 

\subfigure{\includegraphics[clip, trim=17 24 5 22, angle=270,width=.32\linewidth]{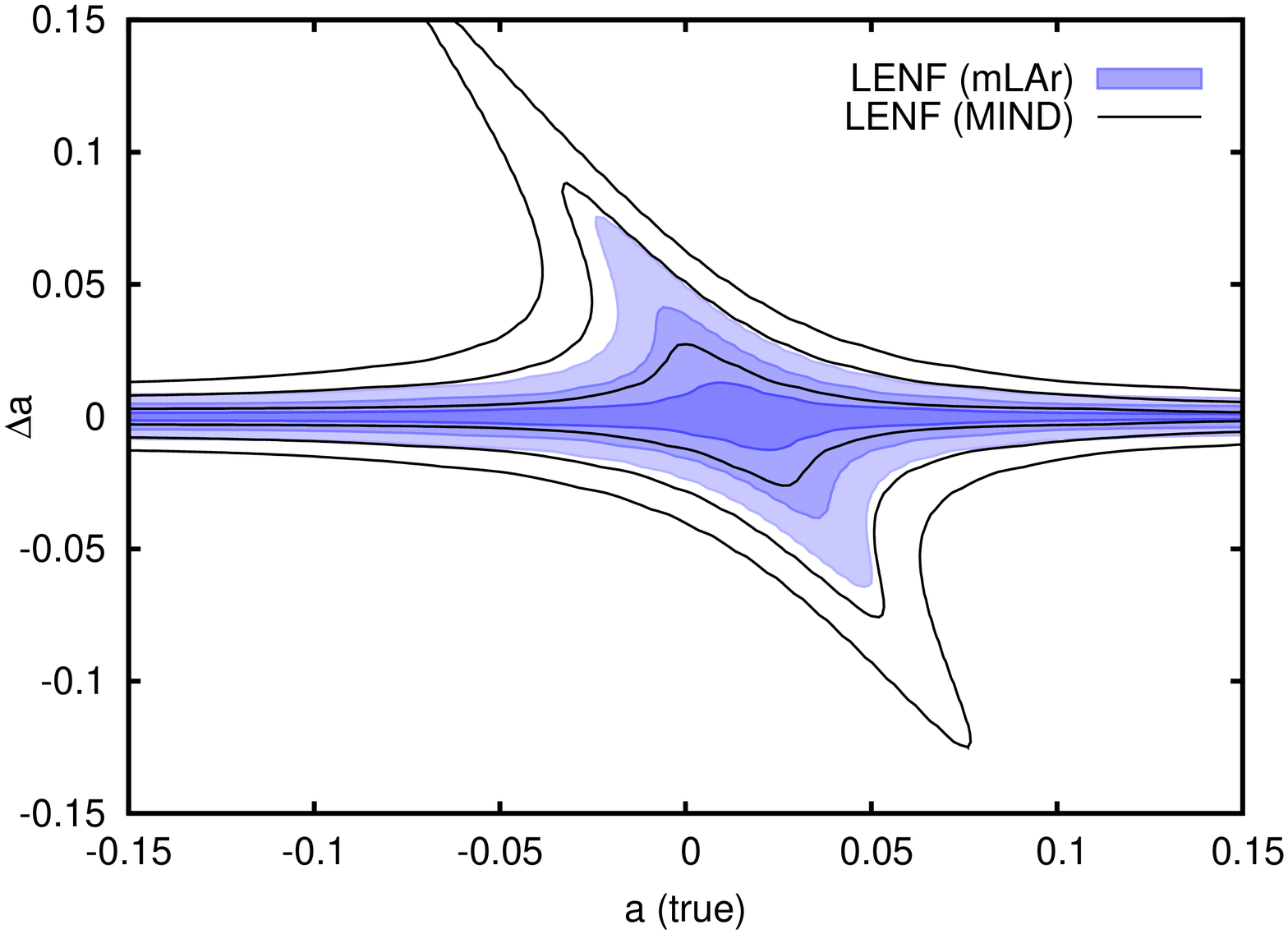}} 
\subfigure{\includegraphics[clip, trim=17 24 5 22, angle=270,width=.32\linewidth]{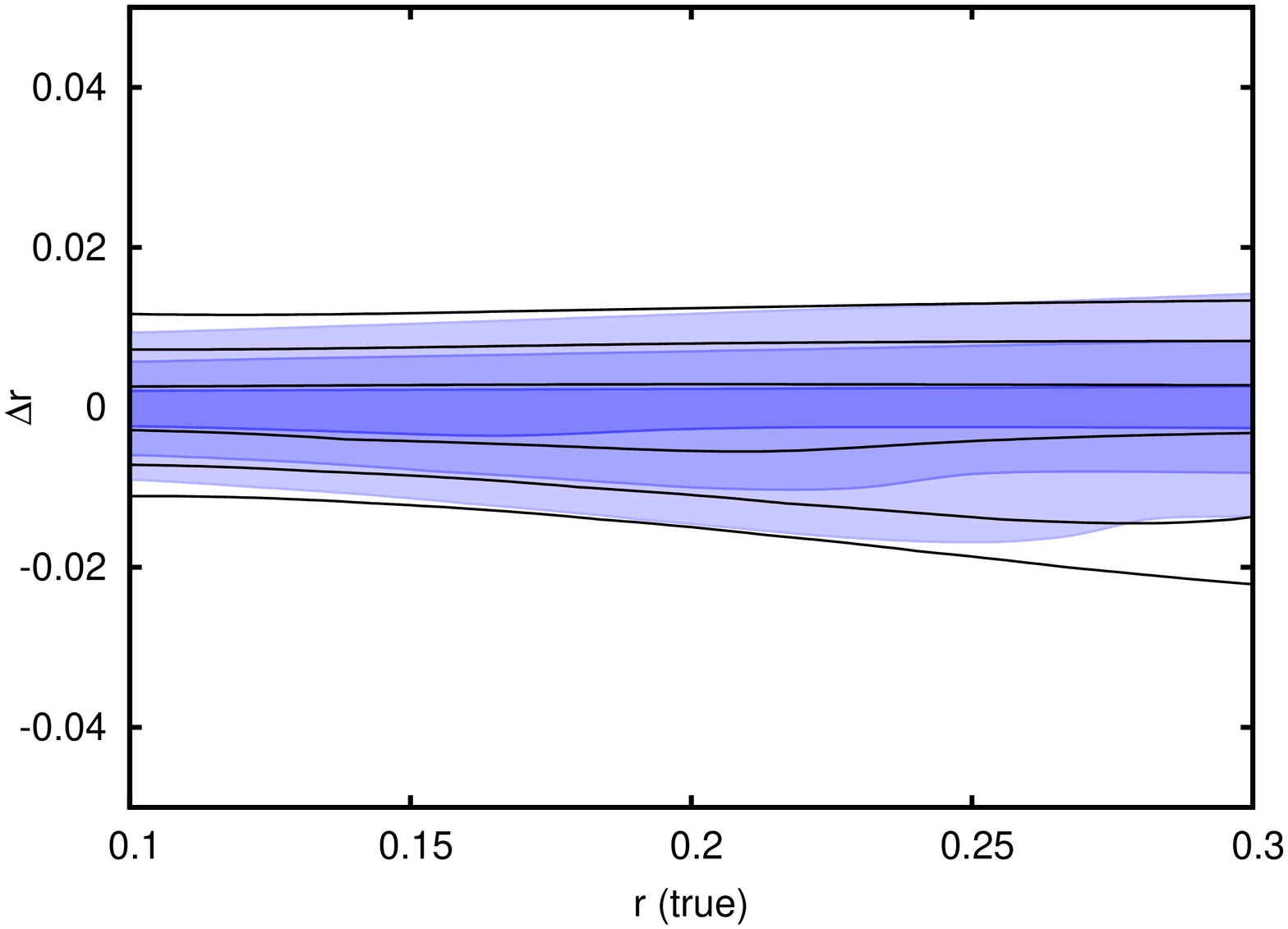}} 
\subfigure{\includegraphics[clip, trim=17 24 5 22, angle=270,width=.32\linewidth]{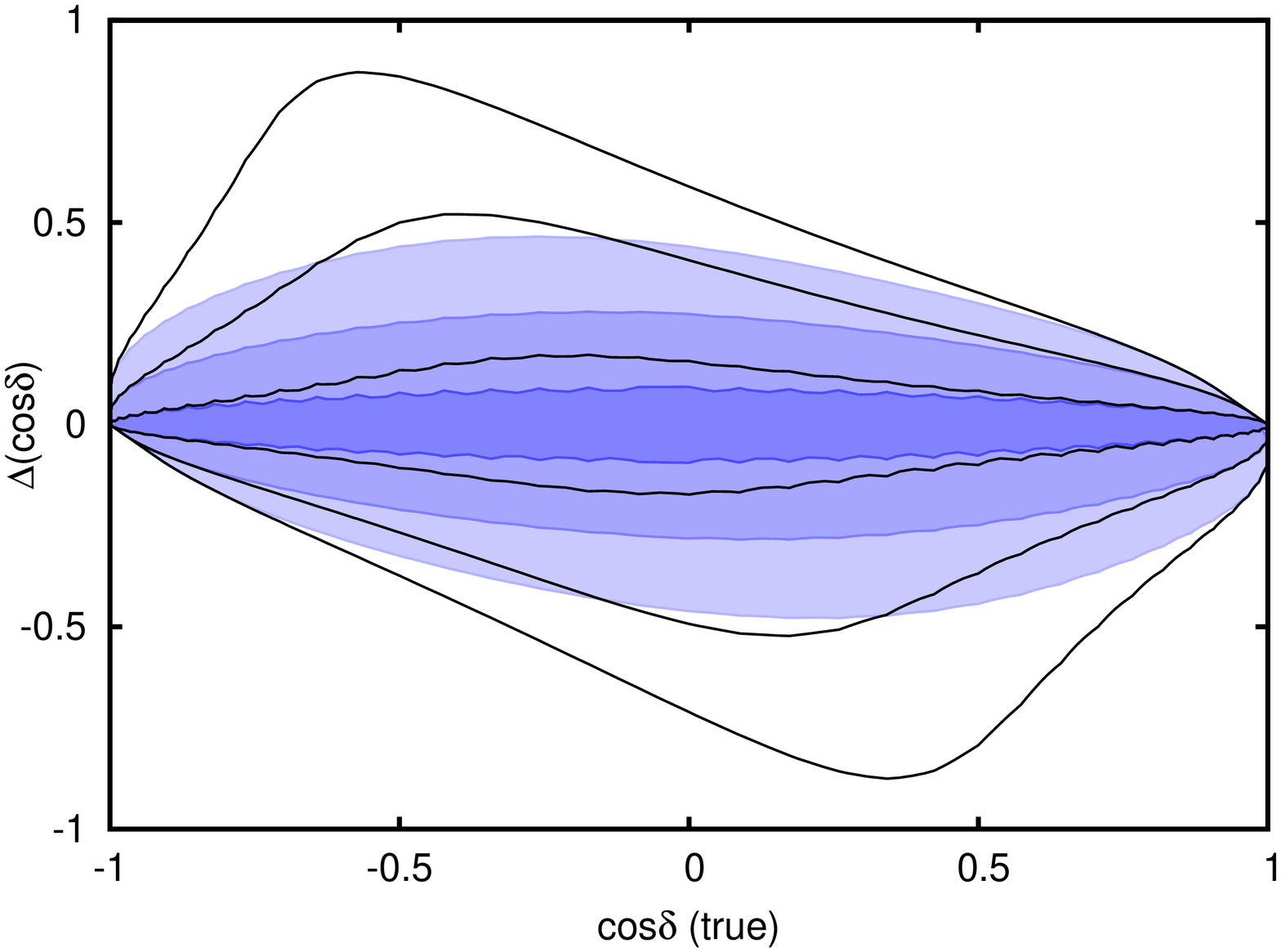}} 

\caption{\label{delAll}The sensitivity of the next-generation facilities to the
$a$, $r$ and $\cos\delta$ parameters. In all of the plots, the shaded regions
progressively show the $1\sigma$, $3\sigma$ and $5\sigma$ regions for the
WBB 70 kton (top row) or the LENF with $50$ kton magnetized LAr
(bottom row), whilst the solid lines are the equivalent envelopes for the WBB
$35$~kton (top row) or the LENF with MIND (bottom row). The leftmost plot
shows the sensitivity to $a$, whilst the central (rightmost) plot shows the
sensitivity to $r$ ($\cos\delta$).} 
\end{figure*}

We have used the GLoBES package~\cite{globes1, *globes2} to perform our
simulations of the LENF and WBB experiments. 
Our model of the WBB design is based on \refref{Coloma:2012wq}, and assumes
$10^{21}$ protons on target per year at $50$~GeV, a baseline distance of
$2300$~km and a $70$~kton ($35$ kton) liquid Argon detector similar to
the GLACIER \cite{Rubbia:2009md} design.  The fluxes for this set-up are taken
from \refref{Longhin:web} (for discussion see \refref{Longhin:2010zz}). We have
assumed a $90\%$ detection efficiency and the backgrounds are taken as arising
from a combination of the contamination of the beam and $0.5\%$ of
neutral-current events at the detector. 
The detector has a low-energy threshold of $100$~MeV with an energy resolution
taken to be a flat $150$~MeV for electrons and $200$~MeV$\times\sqrt{E/\text{GeV}}$ 
for muons.
An error of $5\%$ has been imposed on the signal and background, and a $2\%$
uncertainty on the matter density.
All of our simulations of the LENF design assume $10^{22}$ total useful muon
decays divided equally between $\mu^-$ and $\mu^+$.
The LENF operates with a stored-muon energy of $10$~GeV and a
baseline distance of $2000$~km. These have been shown to be near optimal
choices for large
$\theta_{13}$~\cite{Agarwalla:2010hk,Dighe:2011pa,Ballett:2012rz}. Similar
parameter choices have recently been recommended by the EUROnu Design Study
\cite{Bertolucci:2012fb}, and coincide with the expected specifications of the
International Design Study for the Neutrino Factory \cite{Soler:2011zzb}.
The assumptions in our model of the MIND have
been kindly provided by P.~Soler and R.~Bayes and are based on ongoing work,
evolving from the proposals of \refref{Choubey:2011zz}, which has been
recently reviewed in \refref{Bayes:2013aa}. This model uses migration matrices
to simulate both the appearance and disappearance channels, and considers
backgrounds of charge misidentification, neutral current events and tau
contamination.   
In our model of the mLAr, we have
assumed a threshold energy of $0.5$~GeV and a detection efficiency of $73\%$ at
the lowest energies, rising to $94\%$ at $1$~GeV. The energy resolution is a
flat $10\%$ and the background to the golden channel is taken as $0.1\%$ of the
incident right-sign muons, which models instances of charge misidentification,
and $0.1\%$ of the neutral current events. We have imposed a $2\%$ systematic
uncertainty on both the signal and backgrounds, and a $2\%$ uncertainty on the
matter density. 

The background to the appearance signals caused by $\nu_\tau$ particles
incident on the detectors, which produce electrons and muons by $\tau$ decay,
is known as $\tau$-contamination \cite{Indumathi:2009hg,Dutta:2011mc}. It is
known that this background affects the attainable sensitivity to the
oscillation parameters, causing significant systematic shifts if not properly
taken into account \cite{Donini:2010xk}. The degree with which an experiment
can control the $\tau$-background differs by design. At the LENF, the dominant
$\tau$ particles are right-sign, and only significantly impact the
disappearance channel measurements. Under the assumption that $\cos\delta$ will
introduce the dominant uncertainty in the measurement of sum rules, we can
conclude that the impact of $\tau$-contamination should be slight. For the WBB,
the $\tau$-contamination will affect both appearance and disappearance
channels.  However, the greater kinematic information attainable with LAr
detectors can significantly reduce the impact of this background: a cut-based
analysis on transverse momentum is very effective at removing leptons
originating from $\tau$ decay \cite{Stahl:2012exa}. Therefore, to fairly
implement the $\tau$-contamination effect, we must use information from the
experimental groups working on these detectors. This information is not
available for LAr detectors, and we have chosen to omit the $\tau$-background
at all of the facilities when we are making a direct comparison of performance.
The full implementation of $\tau$-contamination is possible for the LENF with
MIND, and we have checked that there is no significant impact on our
conclusions.

%%%%%%%%%%%%%%%%%%%%%%%%%%%%%%%%%%%%%%%%%%%%%%%%%%%%%%%%%%%%%%%%%%%%%%%%%%%%%%%
%%%%%%%%%%%%%%%%%%%%%%%%%%%%%%%%%%%%%%%%%%%%%%%%%%%%%%%%%%%%%%%%%%%%%%%%%%%%%%%
%%%%%%%%%%%%%%%%%%%%%%%%%%%%%%%%%%%%%%%%%%%%%%%%%%%%%%%%%%%%%%%%%%%%%%%%%%%%%%%
%%%%%%%%%%%%%%%%%%%%%%%%%%%%%%%%%%%%%%%%%%%%%%%%%%%%%%%%%%%%%%%%%%%%%%%%%%%%%%%
%%%%%%%%%%%%%%%%%%%%%%%%%%%%%%%%%%%%%%%%%%%%%%%%%%%%%%%%%%%%%%%%%%%%%%%%%%%%%%%
%%%%%%%%%%%%%%%%%%%%%%%%%%%%%%%%%%%%%%%%%%%%%%%%%%%%%%%%%%%%%%%%%%%%%%%%%%%%%%%

\subsection{\label{sec:precis}Precision for $a$, $r$ and $\cos\delta$}

We start our study by computing the precision with which the next-generation
facilities can individually measure the parameters $a$, $r$ and $\cos\delta$.
An understanding of this precision should give us an indication of the
potential precision towards generic sum rules in these variables, and help us
to identify the dominant uncertainties and functional dependence of such a
measurement.
In the following analysis, we will refer to the parameter values which are used
to generate the simulated data as the \emph{true values} and the parameters
which are extracted by fitting our models to the data as the \emph{fitted
values}. When necessary, true and fitted values will be distinguished by
subscripts \emph{i.e.} $a_\text{T}$ and $a_{\text{F}}$. For each
parameter of interest, we have scanned over a range of true values and then
computed the allowed region (at $1$, $3$ and $5\sigma$) in the fitted value of
this parameter for both experimental set-ups, each with two different
detector options. We marginalize over all of the otherwise unspecified
oscillation parameters in each case. 
The allowed regions are then expressed as a function of the true
parameter value and the difference between the fitted and true values.

The leftmost column in \figref{delAll} shows the sensitivity to $a$ for both
the LENF (bottom row, solid lines for MIND and shaded regions for $50$~kton
magnetized LAr) and the WBB (top row, solid lines for $35$~kton
and shaded regions for $70$~kton detectors). For large values of
$a_\text{T}$, we find the magnitude of $\Delta a \equiv a_{\text{F}} -
a_{\text{T}}$ to be between $0.005$ and $0.015$ at $3\sigma$ for the LENF,
whilst the WBB has worse performance with a range of between $0.014$ and
$0.021$. The attainable precision worsens notably for both experiments
around $|a_{\text{T}}| \lesssim 0.05$, where $\Delta a$ can become potentially
as high as $0.041$ ($0.089$) for the LENF with magnetized LAr (MIND) and
$0.117$ ($0.210)$ for the WBB with $35$~kton ($70$~kton) LAr. This increase is
due to the presence of a degeneracy. For a given value of $a_{\text{T}}$, we
get two reasonably good solutions for the fit $a_{\text{F}} \approx \pm
a_{\text{T}}$: a manifestation of the $\theta_{23}$ octant degeneracy
\cite{Barger:2001yr}.  This is not an exact degeneracy of the $3$-neutrino
oscillation probability, and the ambiguity only appears for the smallest
deviations from $\theta_{23}$-maximality. For all values of $a_\text{T}$, WBB
performs worse than the LENF, and for both facilities, the optimistic detectors
perform better than the more conservative ones. However, if we focus on the
best-fit values for $a$ given by recent global fits, at around
$a=-0.09$~\cite{GonzalezGarcia:2012sz}, the discrepancy between the four
experimental designs considered here is small, with a difference of around
$\pm0.003$ at $1\sigma$, less than $3\%$ of the best-fit value of $a$.

In the middle column of \figref{delAll}, we have computed the sensitivity of
the LENF and WBB to the parameter $r$. Over the region of $r_\text{T}$ that is
phenomenologically interesting, this sensitivity is relatively constant at
about $0.007$ ($0.025$) for the LENF (WBB) at $3\sigma$. There is a slight
broadening of the allowed region towards larger values of $r$; an effect which
is less marked for weaker confidence levels. Once again, we see that LENF
uniformly out-performs WBB. The discrepancy is particularly marked at $5\sigma$
where the WBB allowed region is around $3.5$ times broader than the
corresponding region for the LENF. In recent work on the precision of
next-generation facilities, it has been shown~\cite{Coloma:2012wq} that only
the LENF will be able to surpass the precision on $\theta_{13}$ that is
expected to be attained by the current generation of reactor experiments.
However, the improvement in precision possible with the LENF is rather small,
at around $1\%$, and effectively, the constraints on $\theta_{13}$ will be set
by the reactor experiments alone~\cite{Wang:2013aa}. For this reason,
the observed discrepancy in precision for $r$ between the LENF and WBB is only
expected to influence the ability of the experiments to place individual
constraints on sum rules, and should not influence constraints extracted from
global analyses of the oscillation data.

The rightmost column of \figref{delAll} shows the expected sensitivity to
$\cos\delta$ for the LENF and WBB.  This measurement has a $3\sigma$ precision
at its widest point of $0.28$ ($0.53$) for the LENF with magnetized LAr
(MIND) and $0.65$ ($0.89$) for the WBB with $35$~kton ($70$~kton) LAr. This
decreases dramatically for the extreme points of the spectrum, where the true
value of $\cos\delta$ approaches $\pm1$ and the uncertainty becomes very small
for the LENF, whilst reducing but remaining sizable at higher significances for
WBB. We see that the LENF performs significantly better at this measurement
than WBB: at $5\sigma$, even the WBB with $70$~kton LAr offers little
discriminatory power, with a region that almost covers the whole parameter
space, while the LENF offers a reasonable precision which becomes excellent for
large values of $|\cos\delta|$. The boundaries of the allowed regions at low
significance can be approximated analytically as ellipses: this can be seen by
considering a uniform precision on $\delta$ itself, $\Delta\delta = \epsilon$,
which implies $\Delta(\cos\delta) \equiv \cos\delta_F - \cos\delta_T =
-\epsilon \sin\delta_T + \mathcal{O}(\epsilon^2)$. The coordinates
$(-\epsilon\sin\delta, \cos\delta)$ provide a parametric description of the
ellipse. The assumption of approximately uniform precision in $\delta$ is
consistent with the simulations performed in \refref{Coloma:2012wq} where
$\Delta\delta \approx 5^\circ \pm 2^\circ$ for all $\delta_\text{T}$. The
deviations from ellipticity can be explained by assuming a variable precision
on $\delta$ as shown in \refref{Coloma:2012wq}.  
Generally, $\cos\delta$ is considerably harder to constrain than $r$ and $a$.
As such, it is expected to introduce a significant uncertainty and should be
the dominant limiting factor in the possible constraints on sum rules of the
type shown in \equaref{general-sum rule}. 
However, we must remember that the measurements in this section have focused on
a single parameter at a time, and therefore their results can not be simply
combined to understand the precision on a sum rule. Measurements of parameter
combinations will in general introduce correlations which may strongly
influence the precision, as we will see in the next section.

%%%%%%%%%%%%%

\begin{figure}[t] \centering
\includegraphics[clip,trim=120 110 90 155,width=0.6\linewidth, angle=270]{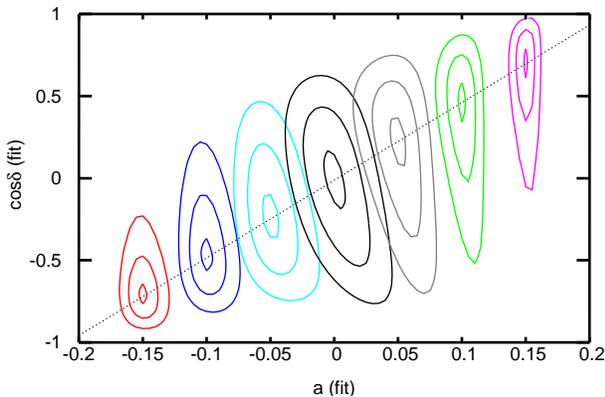} 
\caption{\label{potato}The joint determination of $a$ and $\cos\delta$
for seven sets of true values which obey the relation $a=r\cos\delta$, assuming
the LENF with MIND and including $\tau$ contamination effects. The dashed line
shows the sum rule, and the concentric solid lines indicate the boundary of the
$1$, $3$ and $5\sigma$ allowed intervals for the true values of $a$ and
$\cos\delta$ at their center.} 
\end{figure}

\begin{figure*}[t] \centering

\subfigure{\includegraphics[clip, trim = 15 10 5 25, angle=270,width=0.49\linewidth]{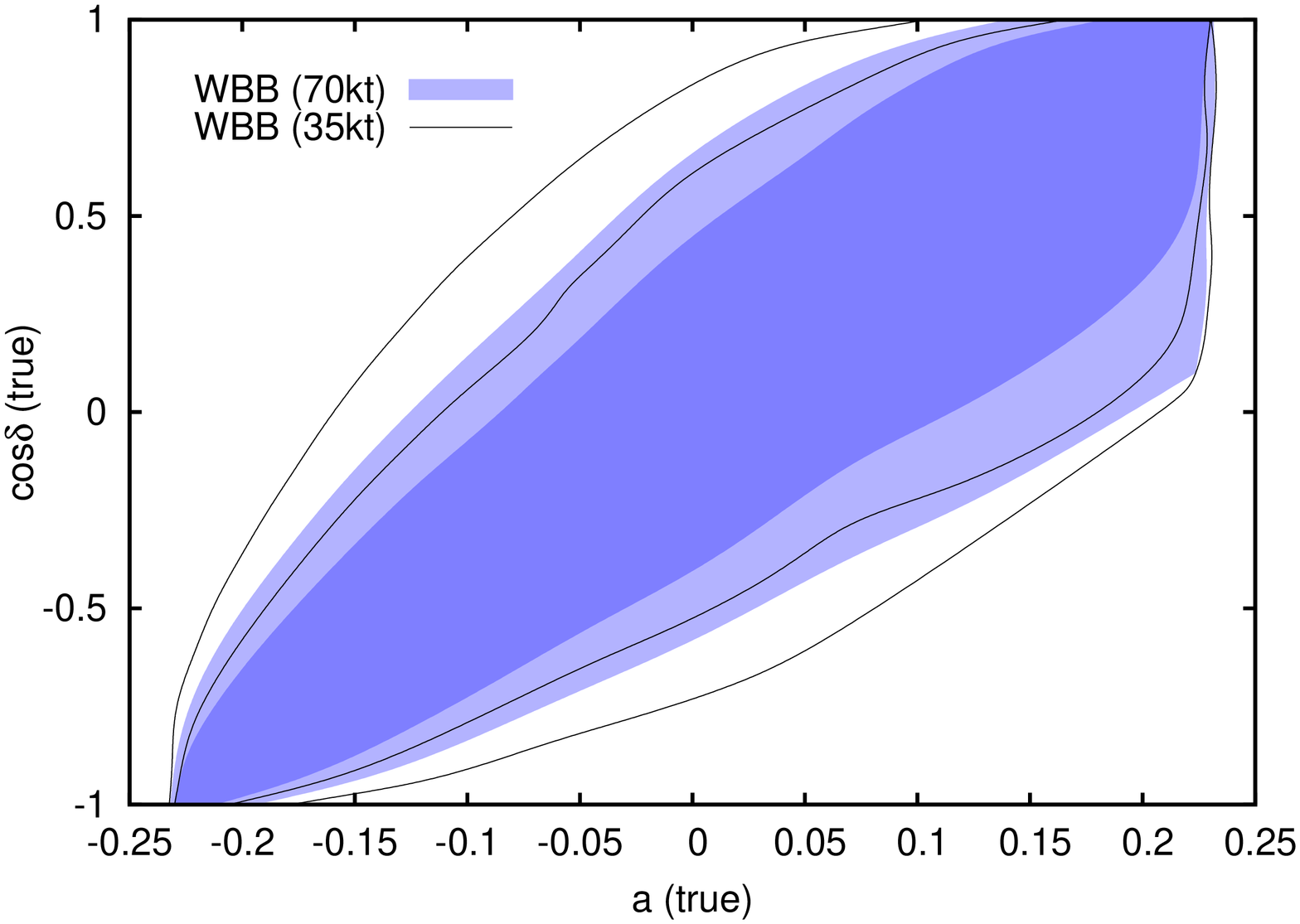}}
\hfill
\subfigure{\includegraphics[clip, trim = 15 10 5 25, angle=270,width=0.49\linewidth]{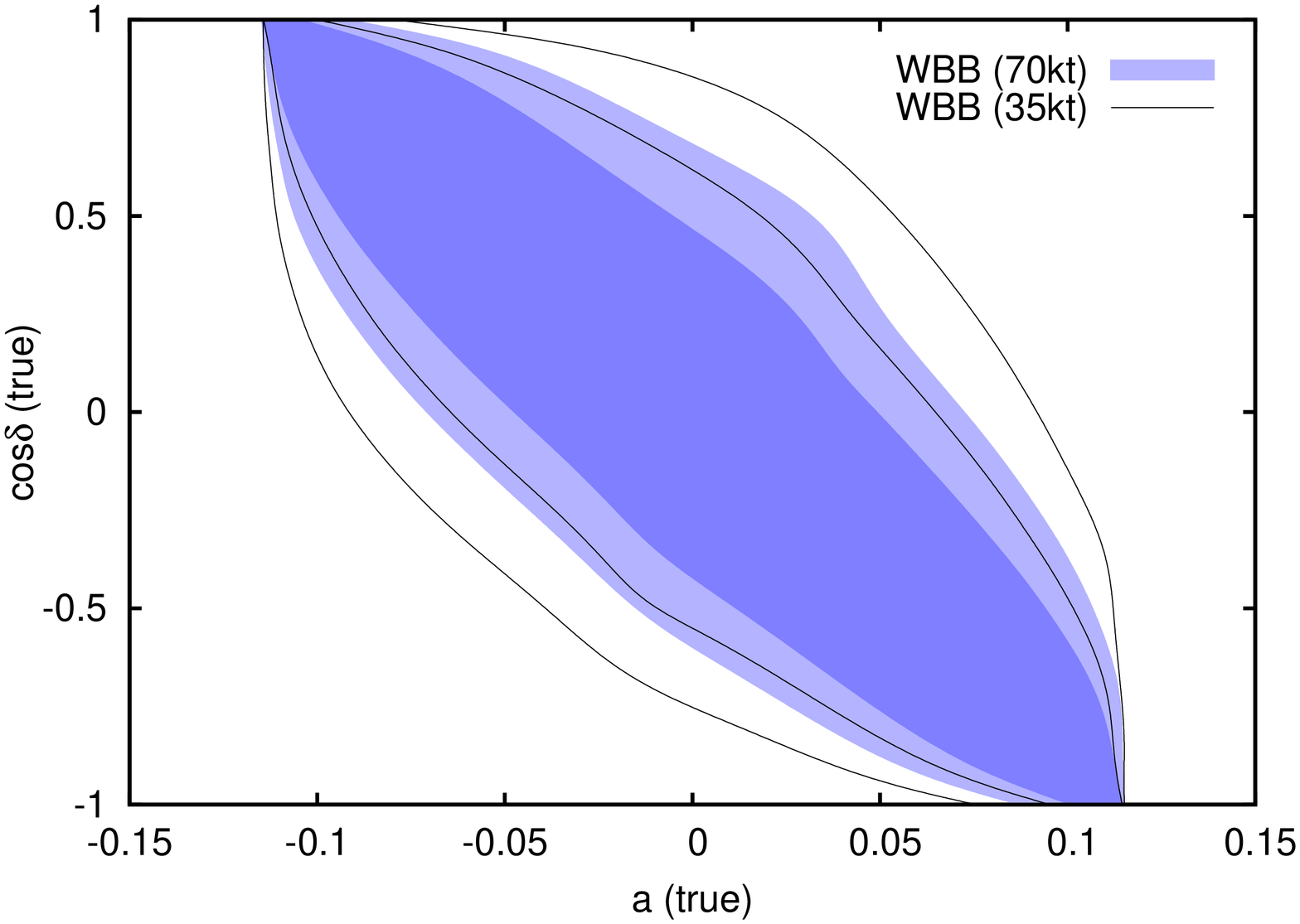}}

\subfigure{\includegraphics[clip, trim = 15 10 5 25, angle=270,width=0.49\linewidth]{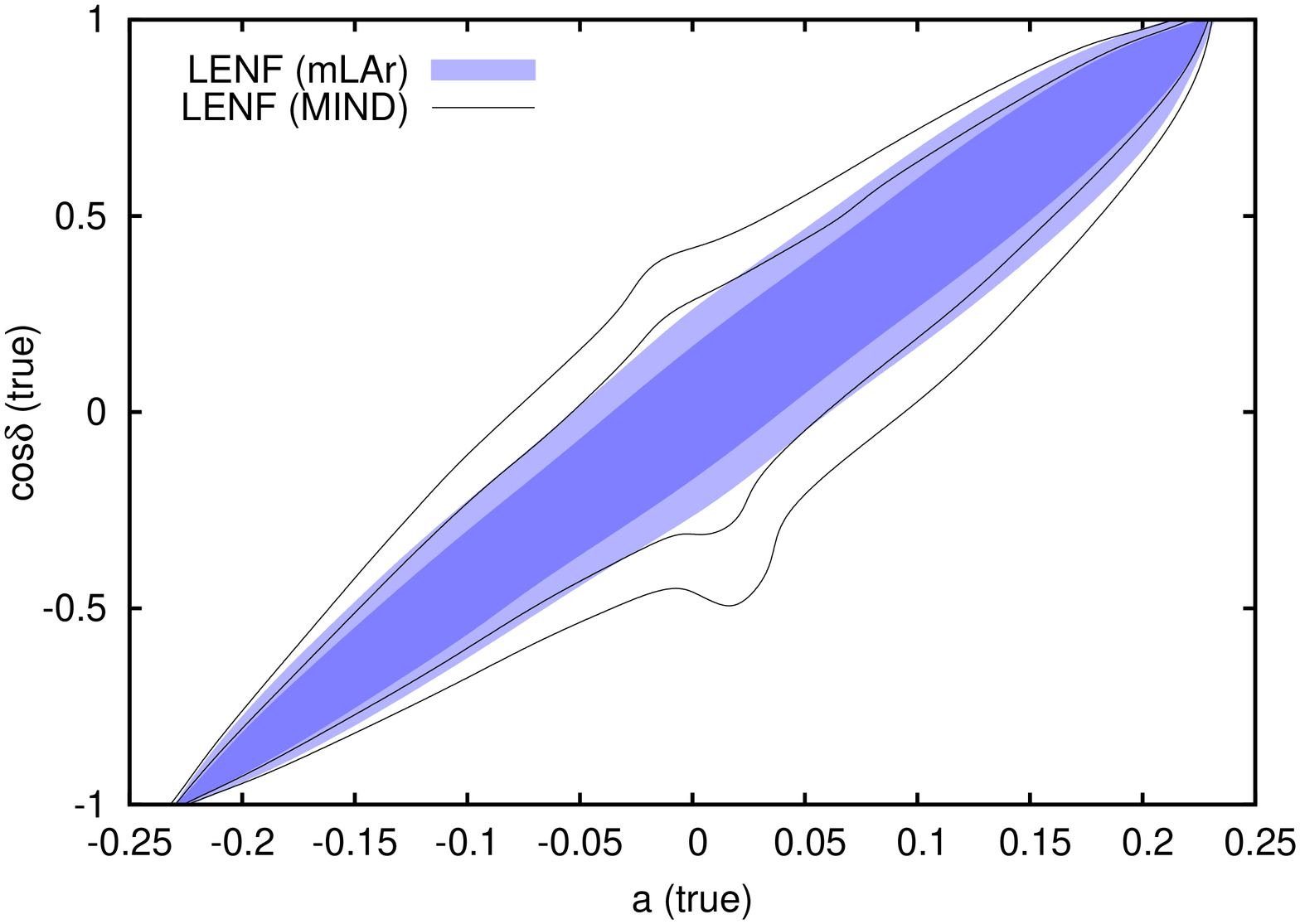}}
\hfill
\subfigure{\includegraphics[clip, trim = 15 10 5 25, angle=270,width=0.49\linewidth]{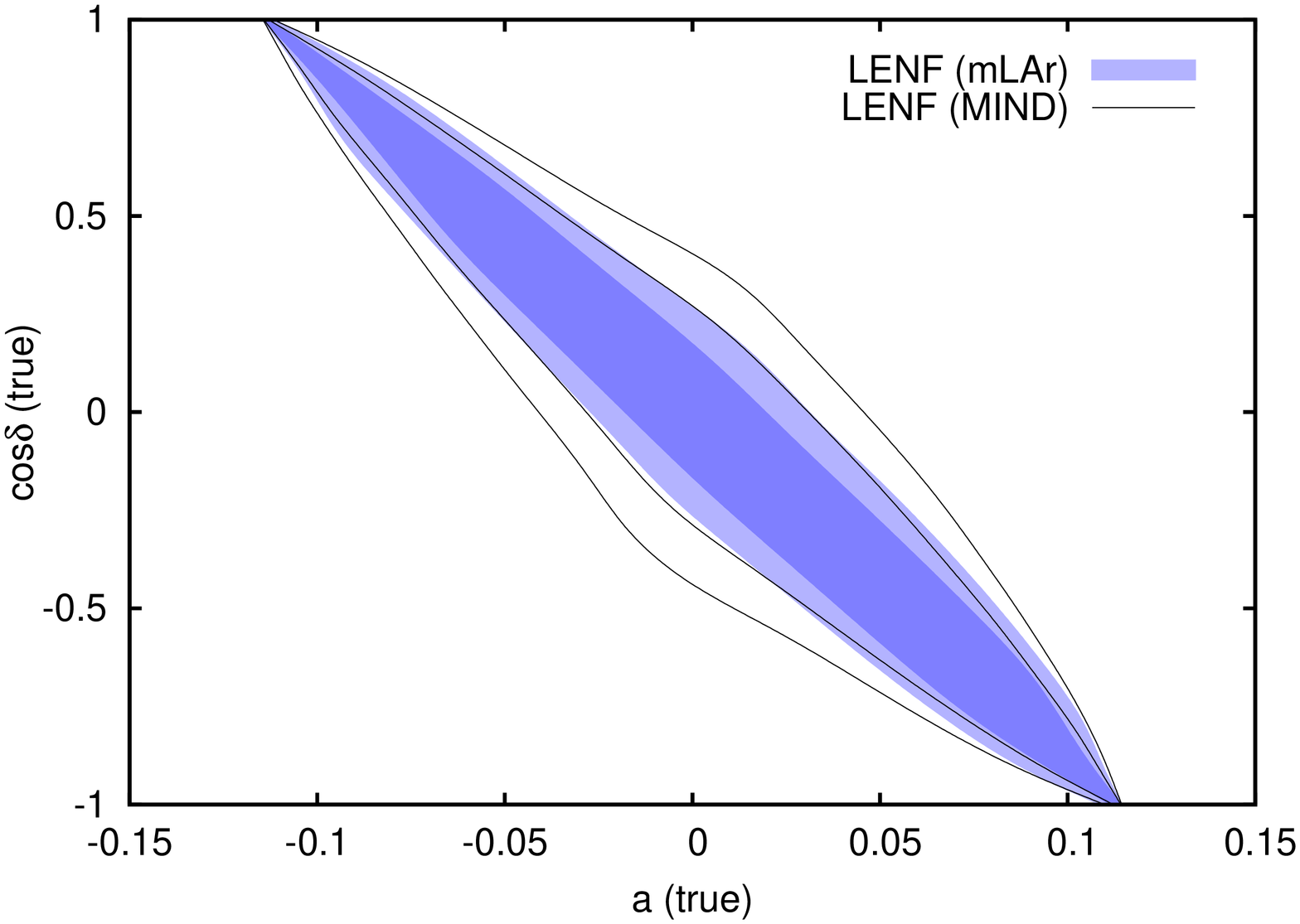}}

\caption{\label{graph1}The left (right) column shows the ability to exclude
models with $\lambda=1$ ($\lambda=-0.5$) as a function of the true parameters.
The plots show the $2\sigma$ and $3\sigma$ allowed regions for
the WBB (top row) and the LENF (bottom row). A point lying
outside of the contours indicates that the model can be excluded by that given
experiment for those true parameter values. }  
 \end{figure*}

%%%%%%%%%%%%%%%%%%%%%%%%%%%%%%%%%%%%%%%%%%%%%%%%%%%%%%%%%%%%%%%%%%%%%%%%%%%%%%%
%%%%%%%%%%%%%%%%%%%%%%%%%%%%%%%%%%%%%%%%%%%%%%%%%%%%%%%%%%%%%%%%%%%%%%%%%%%%%%%
%%%%%%%%%%%%%%%%%%%%%%%%%%%%%%%%%%%%%%%%%%%%%%%%%%%%%%%%%%%%%%%%%%%%%%%%%%%%%%%
%%%%%%%%%%%%%%%%%%%%%%%%%%%%%%%%%%%%%%%%%%%%%%%%%%%%%%%%%%%%%%%%%%%%%%%%%%%%%%%
%%%%%%%%%%%%%%%%%%%%%%%%%%%%%%%%%%%%%%%%%%%%%%%%%%%%%%%%%%%%%%%%%%%%%%%%%%%%%%%
%%%%%%%%%%%%%%%%%%%%%%%%%%%%%%%%%%%%%%%%%%%%%%%%%%%%%%%%%%%%%%%%%%%%%%%%%%%%%%%

\subsection{\label{sec:potatoes}Joint parameter determination}

As a first step to understanding the correlations between the measurements of
oscillation parameters, we have studied how accurately the parameters $a$ and
$\cos\delta$ can be jointly determined for true values which obey a given sum
rule. The correlations between the two parameters will show how strongly the
true value of one parameter influences the determination of the other. In
\figref{potato}, we have computed the joint determination of the parameters
$\cos\delta_\text{F}$ and $a_\text{F}$ for a selection of sets of true
parameters which obey the sum rule $a_\text{T}=r_\text{T}\cos\delta_\text{T}$,
with $r_\text{T}$ fixed at its best fit value derived from global fits of
neutrino oscillation data. This simulation uses the LENF with MIND
experiment, and incorporates the $\tau$-background which is known to impact the
attainable precision on $a$. This plot gives us an indication of the severity
of correlations between these two parameters.  We see that there is some
correlation: the allowed intervals for $\cos\delta$ depend on the true  values
of $a$. The width of the allowed regions in both parameters decreases for large
absolute values of $|a|$ and $|\cos\delta|$, and this behaviour can be
understood by comparing it with the results of \secref{sec:precis}, where the
precision to both $a$ and $\cos\delta$ becomes worse near the origin.  

The joint parameter determination plot can give us an indication of how well we
can measure the parameters $a$ and $\cos\delta$ if the sum rule is true.  In
this plot we have assumed that the true parameters obey the sum rule
$a=r\cos\delta$, indicated by the dashed line with $r$ set to the best
fit value obtained in the global analysis of neutrino oscillation data,  and
we have marginalized over all parameters other than 
$a_\text{F}$ and $\cos\delta_\text{F}$. The solutions found in the allowed
regions are not required to obey the sum rule.  For example, although there are
plenty of solutions around the origin for $a_\text{T}=\cos\delta_\text{T}=0$,
the parameter $r_\text{F}$ is allowed to vary in the marginalization and can
take any reasonable value, meaning that the final solution rarely satisfies
$a=r\cos\delta$. If we are interested in excluding the sum rule without
assuming its validity, we must ask a slightly different question: for a general
set of true parameter values, which sets of parameters obeying a hypothesised
sum rule can be excluded. We will address this question in the next section.

%%%%%%%%%%%%%%%%%%%%%%%%%%%%%%%%%%%%%%%%%%%%%%%%%%%%%%%%%%%%%%%%%%%%%%%%%%%%%%%
%%%%%%%%%%%%%%%%%%%%%%%%%%%%%%%%%%%%%%%%%%%%%%%%%%%%%%%%%%%%%%%%%%%%%%%%%%%%%%%
%%%%%%%%%%%%%%%%%%%%%%%%%%%%%%%%%%%%%%%%%%%%%%%%%%%%%%%%%%%%%%%%%%%%%%%%%%%%%%%
%%%%%%%%%%%%%%%%%%%%%%%%%%%%%%%%%%%%%%%%%%%%%%%%%%%%%%%%%%%%%%%%%%%%%%%%%%%%%%%
%%%%%%%%%%%%%%%%%%%%%%%%%%%%%%%%%%%%%%%%%%%%%%%%%%%%%%%%%%%%%%%%%%%%%%%%%%%%%%%
%%%%%%%%%%%%%%%%%%%%%%%%%%%%%%%%%%%%%%%%%%%%%%%%%%%%%%%%%%%%%%%%%%%%%%%%%%%%%%%

\subsection{\label{sec:violations}Excluding sum rules}

The computation of the attainable sensitivity to combinations of oscillation
parameters differs from the discussion of the previous section, due to the
introduction of non-trivial parameter correlations. In this section, we compute
the ability of the LENF and WBB experiments to directly constrain and exclude the
sum rules discussed in \secref{sec:predict}, whilst fully incorporating
these correlations.

We have scanned over a parameter space spanned by the true value of
$\cos\delta$ and the true value of $a$. At each point in this parameter space,
we have found the best fitting set of oscillation parameters which obey a given
sum rule, and plotted the corresponding value of $\Delta\chi^2$. Once this
value exceeds a chosen significance threshold (for example, $2\sigma$ and
$3\sigma$ in \figref{graph1}), we can consider that sum rule excluded: there
are no sets of parameters which obey that sum rule and provide a reasonable fit
to the data. When the true parameter set approximately obeys the sum rule in
question, we get a good fit, and the width of the surrounding allowed region
gives an indication of how sensitive the experiment is to deviations from the
sum rule. 
Technically, this search has been implemented by using a modified form of the
$\Delta\chi^2$ statistic. We have extended the $\Delta\chi^2$ to include an
additional prior which enforces the sum rule on the set of fitted parameters,
\[\Delta\chi^2 \supset \left (\frac{a_\text{F}-a_0-\lambda r_\text{F}
\cos\delta_\text{F}}{\sigma}\right)^2, \]
where $\sigma$ is a parameter chosen to be small, ensuring that the sum rule is
held to high precision. This term forces the minimal parameter set to obey the
sum rule, whilst not dictating any of the values of the parameters themselves.

We have focused our analysis on the two simplest sum rules $\lambda=1$ and
$\lambda=-\frac{1}{2}$ both with $a_0=0$. This is to illustrate the type of
constraints that can be placed on parameter correlations in the PMNS matrix,
but our approach can be easily generalised to include other types of
correlations, beyond the atmospheric sum rules discussed so far. The plots of
the left-hand (right-hand) panel on the bottom row of \figref{graph1} show the
allowed regions for $\lambda=1$ ($\lambda=-0.5$) for the LENF with magnetized
LAr detector (shaded regions) and MIND (contour lines). We see that the largest
allowed region, and therefore the hardest point to exclude the sum rule, is
when $\cos\delta_\text{T} \approx a_\text{T} \approx 0$.  Whilst the best
sensitivity is generally found at large values of
$\left|\cos\delta_\text{T}\right|$. As expected, this behaviour is largely
inherited from the sensitivity to $\cos\delta$; however, around the origin we
see a novel feature associated with solutions of the type $a=0$ and
$\cos\delta=0$. For any hypothetical sum rule of the type $a=\lambda
r\cos\delta$, a trivial solution can be found for
$a_\text{F}=\cos\delta_\text{F} =0$. At this point, the ability to constrain
both $a$ and $\cos\delta$ is weakened, and we find that regardless of the
relationship between the true parameters, provided they are sufficiently close
to the origin, we can use this solution to describe the data and satisfy the
sum rule. This leads to the lobes around the origin, which are visible
particularly for the LENF with MIND (the improved sensitivity to $a$ of the
LENF with mLAr mitigates the impact of these solutions). The mLAr detector
allows for the sum rule to be excluded over a larger region of parameter space:
the $2\sigma$ allowed region for the mLAr is contained completely inside the
$2\sigma$ region for the MIND detector.  At the widest points, the allowed
regions for $\cos\delta$ cover around $24\%$ ($42\%$) of the parameter space
for $\cos\delta$ for the LENF with mLAr (MIND) at $3\sigma$. 
On the top row of \figref{graph1}, we show the equivalent regions for the WBB
with $35$~kton and $70$~kton LAr. These follow the same shape, inherited from
the uncertainties in measurement of $\cos\delta$. In this case, the uncertainty
in $\cos\delta$ is large enough to subsume the lobed solution regions found for
the LENF. The WBB is unable to constrain the parameter $\cos\delta$ to the same
extent as the LENF, and we see that the allowed region for the sum rules are
correspondingly much larger. At its widest point, the WBB with $70$~kton
($35$~kton) LAr has an allowed region for $\cos\delta$ which covers $56\%$
($81\%$) of the parameter space at $3\sigma$. For both LENF and WBB, excluding
models over even $50\%$ of the parameter space would be an interesting result;
however, we have seen that these measurements are challenging, and the more
optimistic facilities are required to make significant advances. 

%%%%%%%%%%%

\begin{figure*}[t] \centering
\subfigure[~$\lambda_\text{F}=-\frac{1}{2}$]{\includegraphics[clip,trim=17 22 7 40, width=0.35\linewidth,angle=270]{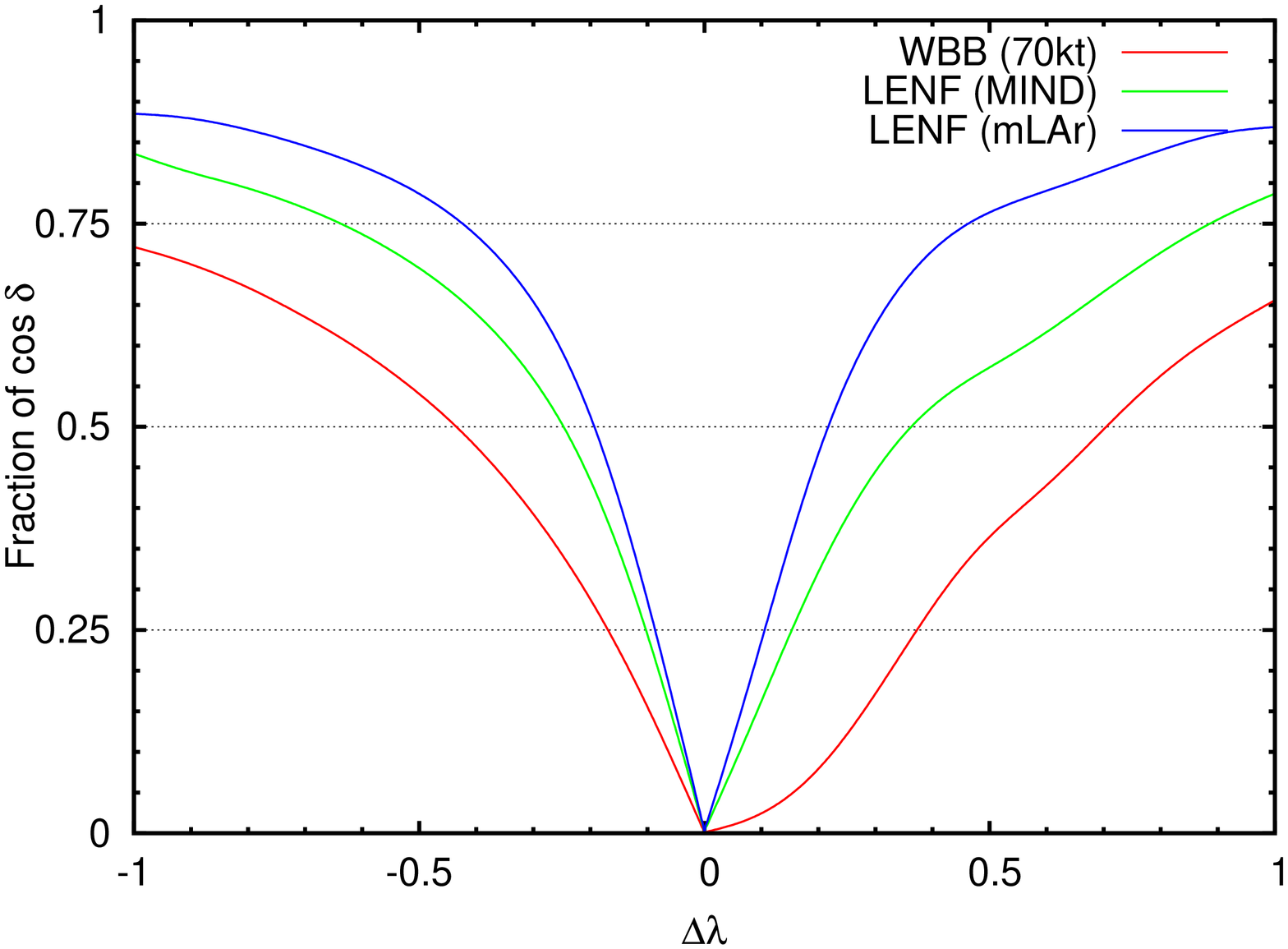}} ~\subfigure[~$\lambda_\text{F}=1$]{\includegraphics[clip, trim=17 10 7 30,width=0.35\linewidth,angle=270]{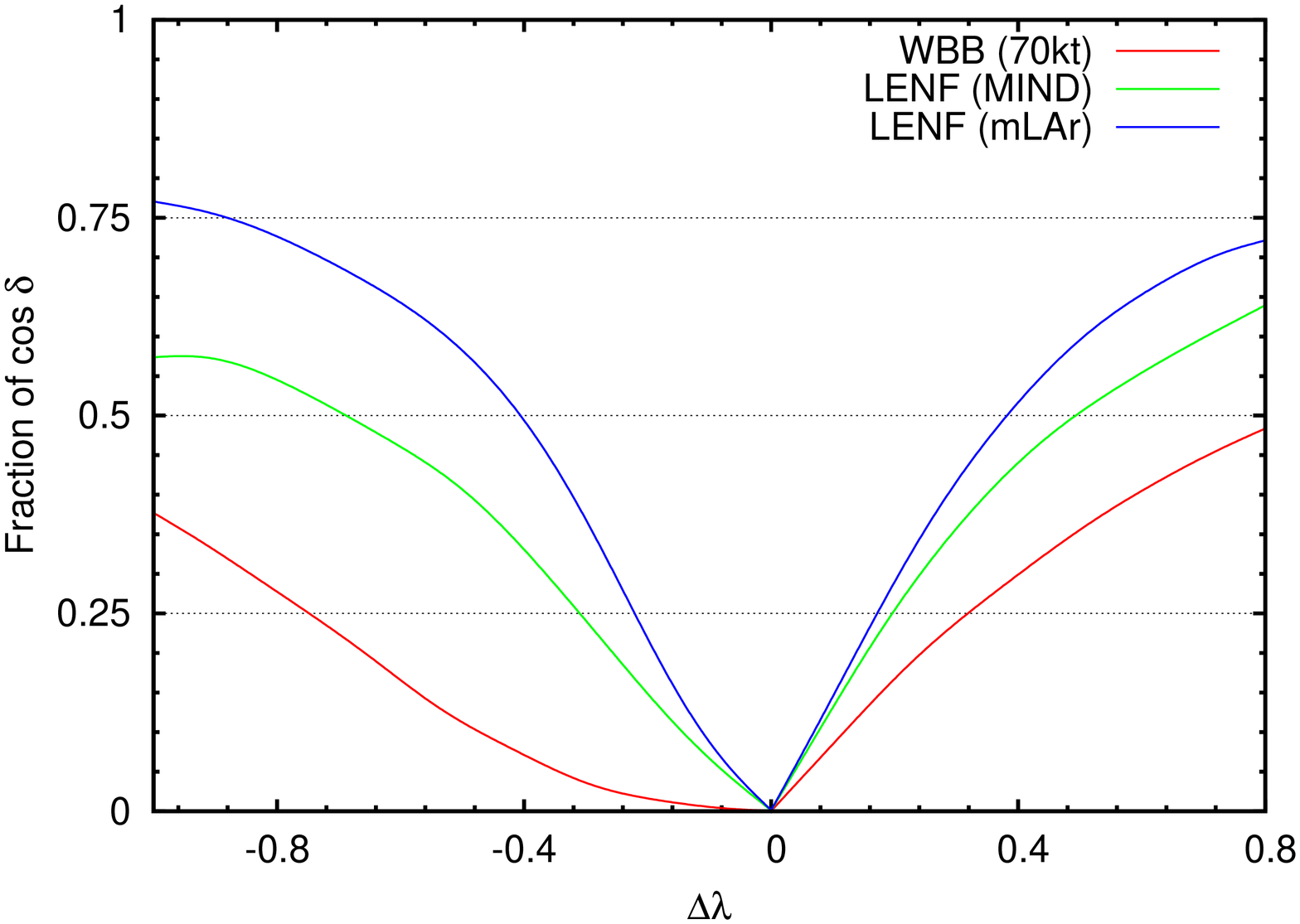} }
\caption{\label{deltalam} The fraction of values of
$\cos\delta_{\text{T}}$ for which the hypothesised value of
$\lambda_\text{F}$ can be excluded at $3\sigma$ assuming different true
values of  $\lambda_\text{T}$. In these plots $\Delta\lambda = \lambda_\text{T}
- \lambda_\text{F}$.}  
 \end{figure*}

%%%%%%%%%%%%%%%%%%%%%%%%%%%%%%%%%%%%%%%%%%%%%%%%%%%%%%%%%%%%%%%%%%%%%%%%%%%%%%%
%%%%%%%%%%%%%%%%%%%%%%%%%%%%%%%%%%%%%%%%%%%%%%%%%%%%%%%%%%%%%%%%%%%%%%%%%%%%%%%
%%%%%%%%%%%%%%%%%%%%%%%%%%%%%%%%%%%%%%%%%%%%%%%%%%%%%%%%%%%%%%%%%%%%%%%%%%%%%%%
%%%%%%%%%%%%%%%%%%%%%%%%%%%%%%%%%%%%%%%%%%%%%%%%%%%%%%%%%%%%%%%%%%%%%%%%%%%%%%%
%%%%%%%%%%%%%%%%%%%%%%%%%%%%%%%%%%%%%%%%%%%%%%%%%%%%%%%%%%%%%%%%%%%%%%%%%%%%%%%
%%%%%%%%%%%%%%%%%%%%%%%%%%%%%%%%%%%%%%%%%%%%%%%%%%%%%%%%%%%%%%%%%%%%%%%%%%%%%%%

\subsection{Constraining $\lambda$} Both LENF and WBB will be able to observe
violations of a given sum rule for a significant fraction of parameter space,
especially if $|\cos\delta|$ is large. In the scenario that the true parameter
set appears to agree with some sum rule, it is interesting to see what
constraints we can put on the parameters describing such a rule. In this
section, we consider the ability of the next-generation oscillation experiments
to distinguish between models with similar $\lambda$ parameters, introduced in
\equaref{general-sum rule}. Our interest here is in illustrating the possible
constraints that can be posed by a next-generation oscillation experiment, and
as such we will restrict our attention to some specific cases; however, the
analysis of this section could be simply extended to address other classes of
models.

We consider general relations of the type $a=\lambda r\cos\delta$, with
continuous ranges of $\lambda$ in the neighbourhoods of the special values
$\lambda=1$ and $\lambda=-1/2$. The plots in \figref{deltalam} show how well
a hypothesised value of $\lambda_\text{F}$ can  be excluded as a function of
$\Delta\lambda\equiv \lambda_\text{T}-\lambda_\text{F}$.
As the lines of parameters which obey sum rules with $a_0=0$ intersect for
$\cos\delta_\text{T}=0$, we will always be able to find true parameter
values close to this value of $\cos\delta_\text{T}$ which satisfy any pair of 
sum rules. Therefore, it is impossible to distinguish two similar models in
all possible cases, and instead we must assess this ability by degree. 
In order to measure the degree of distinguishability at different facilities,
we have plotted a continuous parameter which gives the fraction of values of
$\cos\delta_\text{T}$ for which we can exclude the hypothesis
$\lambda=\lambda_\text{F}$ at $3\sigma$. 
The corresponding fraction of distinguishability for the hypothesis
$\lambda_\text{F}=-1/2$ ($\lambda_\text{F}=1$) as a function of
$\Delta\lambda$ is shown in the left (right) panel of \figref{deltalam}.
If we choose our threshold to be $50\%$ of all possible values of
  $\cos\delta_\text{T}$, the LENF with mLAr can distinguish between sum rules
of the type $\lambda\approx-1/2$ which deviate by $\left|\Delta\lambda\right|
\approx 0.2$. If we instead use a MIND, this region increases to
$\left|\Delta\lambda\right| \approx 0.3$, whilst the WBB superbeam with a
detector of $70$~kton is closer to $\left|\Delta\lambda\right| \approx
0.7$. For sum rules with $\lambda\approx 1$ the size of these deviations
approximately doubles. 

For the models presented in \secref{sec:sum rules}, which cluster around
$\lambda=1$ or $\lambda=-0.5$, the values of $\lambda$ differ by around
$\pm0.1$.  The ability to separate these candidate models experimentally is
clearly dependent on the true value of $\cos\delta$; however, the LENF with mLAr
can make this discrimination for about $25\%$ of the values $\delta$ at $3\sigma$.
This will be a very challenging measurement and is unlikely to be
feasible in the next-generation of oscillation experiments unless an aggressive
strategy is adopted.  

%%%%%%%%%%%%%%%%%%%%%%%%%%%%%%%%%%%%%%%%%%%%%%%%%%%%%%%%%%%%%%%%%%%%%%%%%%%%%%%
%%%%%%%%%%%%%%%%%%%%%%%%%%%%%%%%%%%%%%%%%%%%%%%%%%%%%%%%%%%%%%%%%%%%%%%%%%%%%%%
%%%%%%%%%%%%%%%%%%%%%%%%%%%%%%%%%%%%%%%%%%%%%%%%%%%%%%%%%%%%%%%%%%%%%%%%%%%%%%%
%%%%%%%%%%%%%%%%%%%%%%%%%%%%%%%%%%%%%%%%%%%%%%%%%%%%%%%%%%%%%%%%%%%%%%%%%%%%%%%
%%%%%%%%%%%%%%%%%%%%%%%%%%%%%%%%%%%%%%%%%%%%%%%%%%%%%%%%%%%%%%%%%%%%%%%%%%%%%%%
%%%%%%%%%%%%%%%%%%%%%%%%%%%%%%%%%%%%%%%%%%%%%%%%%%%%%%%%%%%%%%%%%%%%%%%%%%%%%%%

\section{\label{sec:conclusions}Conclusions}

Next-generation neutrino oscillation facilities are not only necessary to
resolve the traditional questions about the PMNS matrix, but will also lead the
way in a new programme of precision neutrino flavour physics. Over the years,
many attempts have been made to understand the origin of flavour. One popular
approach is to invoke a symmetry to explain the pattern of mixing angles that
have been discovered experimentally in the PMNS matrix: an idea which has met
with great success and generated a large number of candidate models. Thanks to
the precision that is expected at the next-generation oscillation facilities,
it will soon be possible to put these theories to the test. 

A predictive model of flavour will generally introduce correlations amongst the
parameters of the Yukawa sector. The linearized expressions of these
correlations are called sum rules, and testing them is a direct way to confirm
or exclude a given model. In this paper, we have studied how correlations
of the type given in \equaref{general-sum rule} will be constrained by current
and future oscillation experiments.  We have seen that, when viewed as
predictions for $\cos\delta$, these sum rules are constrained by their
consistency with the current data, and although all of the models that we have
investigated have some region of applicability, some models may become quite
constrained in the near future. The major difficulty in constraining the sum
rules found in \secref{sec:sum rules}, is the absence of information on the
parameter $\cos\delta$, and we must look to the next generation of oscillation
experiments to provide this. 
We have studied the ability of two candidate next-generation neutrino
oscillation experiments, a low-energy neutrino factory and a wide-band
superbeam, to constrain these correlations. To illustrate the general
constraints that these experiments can place on flavour effects, we have chosen
to focus our attention on sum rules with the form $a=a_0+\lambda r \cos\delta$,
and specifically on the choices $\lambda=1$ and $\lambda=-0.5$.  These have
arisen previously in the literature, and we have shown in \secref{sec:sum
rules} that these two special values appear to well characterize a large class
of models.  We have seen that violations of these sum rules will be readily
testable at the LENF and WBB: the WBB with $70$~kton ($35$~kton) LAr is
expected to be able to exclude the relation $a=r\cos\delta$ for at least $44\%$
($19\%$) of the parameter space, whilst the LENF with mLAr (MIND) can
make the same exclusion for at least $76\%$ ($58\%$).  We have also
considered the ability to distinguish between models which predict similar sum
rules with separations in $\lambda$ of only around $\pm0.1$. We have found that
this ability is dependent on the exact value of $\cos\delta$; however, it is
likely that only the LENF with magnetized LAr is precise enough to make
such a distinction at a reasonable statistical significance for $25\%$ of the
parameter space. 

We have shown that correlations amongst the parameters of the PMNS matrix, as
in the atmospheric mixing sum rules considered here, may be tested by
the next generation of neutrino oscillation facilities. These correlations
can be excluded for a significant part of the parameter space,
and constraints can be inferred on the underlying models responsible for them.
This not only highlights the important role of the precision neutrino
physics programme in our search for the origin of flavour, but also the great
advances which are possible in the decades to come.

%%%%%%%%%%%%%%%%%%%%%%%%%%%%%%%%%%%%%%%%%%%%%%%%%%%%%%%%%%%%%%%%%%%%%%%%%%%%%%%
%%%%%%%%%%%%%%%%%%%%%%%%%%%%%%%%%%%%%%%%%%%%%%%%%%%%%%%%%%%%%%%%%%%%%%%%%%%%%%%
%%%%%%%%%%%%%%%%%%%%%%%%%%%%%%%%%%%%%%%%%%%%%%%%%%%%%%%%%%%%%%%%%%%%%%%%%%%%%%%
%%%%%%%%%%%%%%%%%%%%%%%%%%%%%%%%%%%%%%%%%%%%%%%%%%%%%%%%%%%%%%%%%%%%%%%%%%%%%%%
%%%%%%%%%%%%%%%%%%%%%%%%%%%%%%%%%%%%%%%%%%%%%%%%%%%%%%%%%%%%%%%%%%%%%%%%%%%%%%%
%%%%%%%%%%%%%%%%%%%%%%%%%%%%%%%%%%%%%%%%%%%%%%%%%%%%%%%%%%%%%%%%%%%%%%%%%%%%%%%

\vspace{0.5cm}
\section*{Acknowledgements}

We would like to thank Thomas Schwetz-Mangold for kindly providing us with the
global-fit data used in \figref{predict}, and Paul Soler and Ryan Bayes for
assisting in the simulation of the MIND.

The authors acknowledge partial support from the European Union FP7 ITN
INVISIBLES (Marie Curie Actions, PITN-GA-2011-289442). They also thank Galileo
Galilei Institute for Theoretical Physics for its hospitality. 
PB is supported by a U.K. Science and Technology Facilities Council (STFC)
studentship.
SFK acknowledges partial support from the STFC Consolidated ST/J000396/1
and EU ITN grants UNILHC 237920.
SP acknowledges the support of EuCARD (European Coordination for Accelerator
Research and Development), which is co-funded by the European Commission within
the Framework Programme 7 Capacities Specific Programme, under Grant Agreement
number 227579. 
MS acknowledges partial support by the Australian Research Council.

\appendix*
\section{\label{sec:hernandez}sum rules in the Hernandez-Smirnov framework}

In \refref{Hernandez:2012ra} a novel approach was developed for the generation
of correlations between the parameters of the PMNS matrix following earlier
work~\cite{Lam:2008sh,*Lam:2009hn,Ge:2011ih,*Ge:2011qn}. The method assumes the
breaking of a discrete flavour group into two distinct $Z_n$ subgroups which
remain unbroken in either the charged lepton or neutrino sector, whilst broken
in the other.
Based on this construction, the authors of \refref{Hernandez:2012ra} reported a
number of parameter correlations; however, these correlations led to sum rules identical to those reported in previous studies. In this section, we
weaken some of the assumptions made in the derivations of these relations and
generate additional correlations with distinct sum rules.  We refer the
reader to \refref{Hernandez:2012ra} for a detailed discussion of the method for
finding parameter correlations in the ``symmetry building'' approach, and we
will only summarize the steps here, highlighting where we alter the derivation.

The approach in \refref{Hernandez:2012ra} assumes that the grand flavour group
is a von Dyck group, $\mathrm{D}(n,m,p)$. These groups are defined by the
presentation 
\[S^n = T^m = W^p = STW = 1. \]
The generators $S$ and $T$ are assumed to describe residual symmetries of the
Majorana neutrino and charged lepton mass terms, respectively; whilst $W$ is
defined to be the inverse of the product $ST$.
The symmetry of the Majorana neutrino mass term is the Klein group
${Z}_2\!\times\!{Z}_2$, which fixes $n$ to be given by $n=2$. Only one of the
$Z_2$ factors originates from the flavour symmetry and is generated by $S$,
while the other one arises accidentally. If the second $Z_2$ would be embedded
in the group as well, another parameter relation would appear, which fixes the
mixing angles as it has been discussed in \refref{Lam:2008sh,*Lam:2009hn}.  The
choice of $m$ and $p$ remains free; however, the assumption that the unbroken
group is finite restricts these to specific values.\footnote{See
\refref{Hernandez:2012sk,*Hu:2012ei,*Grimus:2013rw,*Lam:2013xs,*Hernandez:2013vya}
for further extensions and generalizations of this approach.} Representing
each choice by the ordered pair $(m,p)$, the choices which lead to finite
groups are exhausted by five special pairs
\[ (3,3),\quad(3,4),\quad(3,5),\quad(4,3),\quad(5,3), \]
and $2$ infinite sequences
\[ (2,N)\quad \text{and}\quad (N,2) \qquad \forall\,N\,\ge2. \]
The former are isomorphic to the groups A$_4$, S$_4$, A$_5$, S$_4$, A$_5$,
respectively. The two infinite sequences lead to dihedral symmetry groups which
do not have irreducible triplet representations and are therefore not
considered any further. 

For a given $(m,p)$, the two generators $S$ and $T$ must be chosen from the
symmetries of the leptonic mass terms, assuming that they are residual
symmetries following the spontaneous breakdown of $G_f$. For this to be the
case, the generators $S$ and $T$ must have at least one unit eigenvalue. This
is necessary for there to exist a VEV alignment that remains invariant under
their action. Under the further assumption that the discrete groups are
subgroups of SU($3$), we find that the symmetry of the diagonalised neutrino
mass matrix must be given by either
\begin{align*}&S'_1 = \left(\begin{matrix} 1 &0&0\\0&-1&0\\0&0&-1
\end{matrix}\right),\quad S'_2= \left(\begin{matrix} -1&0&0\\0&1&0\\0&0&-1
\end{matrix}\right),\\ &\qquad\quad\text{or}\quad S'_3= \left(\begin{matrix}
-1&0&0\\0&-1&0\\0&0&1  \end{matrix}\right).   \end{align*}
Similarly, these constraints imply that the symmetry of the diagonalised
charged lepton mass matrix is given by one of the three order-$m$ generators
\begin{align*} &T'_e = \left(\begin{matrix}1&0&0\\0&e^{\mathrm{i}\frac{2\pi
k}{m}}&0\\0&0&e^{-\mathrm{i}\frac{2\pi k}{m}}\end{matrix}\right), ~~~~ T'_\mu =
\left(\begin{matrix}e^{\mathrm{i}\frac{2\pi
k}{m}}&0&0\\0&1&0\\0&0&e^{-\mathrm{i}\frac{2\pi k}{m}}\end{matrix}\right),\\
&\qquad\quad\text{or}\quad T'_\tau =
\left(\begin{matrix}e^{\mathrm{i}\frac{2\pi
k}{m}}&0&0\\0&e^{-\mathrm{i}\frac{2\pi k}{m}}&0\\0&0&1\end{matrix}\right),
\end{align*}
where $k\in\{n\in{Z}_m\,|\,n{\rm~and~}m{\rm~are~coprime}\}$. Working in the
basis of diagonal charged leptons, we have $T_\alpha=T'_\alpha$ and
$S_i=U_\text{PMNS} S_i'U_\text{PMNS}^\dagger$.

With a choice of generators $T_\alpha$--$S_i$, we can construct~$W$
\[  W^{- 1} = S_iT_\alpha =U_\text{PMNS} S_i' U_\text{PMNS}^\dagger T_\alpha. \]
As in \refref{Hernandez:2012ra}, it is assumed that $W$ has an eigenvalue~$1$,
which can be shown to constrain $\mathrm{Tr}[W]$ to be real. For the three
finite von Dyck groups with a $3$-dimensional irreducible representation,
it can be shown by considerations of the group character tables that this is
in fact a necessary property. From the group presentation, we see that the
remaining eigenvalues must be $p$-th roots of unity, and therefore, we can express 
\begin{equation} \mathrm{Tr}[W] = 1 + 2\cos\left(\frac{2\pi d}{p}\right)\qquad
\text{s.t.}~d \in {Z}_p.  \label{a}\end{equation}
Once we have computed $\mathrm{Tr}[W]$ we have fully specified the constraints on the PMNS
matrix. These fix one of the columns of the PMNS matrix, where the column fixed
corresponds to the choice of generator $S_i$, and the order of the rows on the
choice of $T_\alpha$. In general, the constraints are given by
\begin{align*} \left|U_{\beta i}\right|^2 &=\left|U_{\gamma i}\right|^2 =
\frac{1-\eta}{2},\\ \left|U_{\alpha i}\right|^2 &= \eta, \end{align*}
where $\{\alpha,\beta,\gamma\}=\{e,\mu,\tau\}$, and $\eta$ is defined by
\begin{equation*} \eta = \frac{1+\mathrm{Tr}[W]}{4\sin^2\left(\frac{\pi
k}{m}\right)}.\end{equation*}
Combined with \equaref{a}, this produces an expression for $\eta$ in terms of
$k$ and $d$
\begin{equation} \eta = \frac{\cos^2\left(\frac{\pi
d}{p}\right)}{\sin^2\left(\frac{\pi k}{m}\right)}. \label{eta} \end{equation}
In \refref{Hernandez:2012ra} $k$ is fixed so that $k=1$ and $d$ is not varied
systematically. However, by varying these parameters we can find novel
parameter correlations and, as we will show, can generate sum rules which have
not been previously identified in the literature.

As we have mentioned, the constraints imposed by this method fix the $i$-th
column of the PMNS matrix by symmetry alone. The values of the elements of this
column are given by the choice of $(m,p)$ and the choice of two integers $k$
and $d$. Which column is fixed, and the pattern of values that are imposed, is
governed by a choice of one of 9 possible pairs of generators. Only 4 of these
choices appear interesting phenomenologically: $T_e$--$S_1$, $T_e$--$S_2$,
$T_\mu$--$S_2$ and $T_\tau$--$S_2$.\footnote{The remaining 5 pairs
$T_\alpha$--$S_i$, lead to correlations that cannot be reconciled with the
current phenomenological data for any choice of $(m,p)$, $k$ and $d$.} For
these cases, the resulting constraints can always be expressed by two
relations: the first leads to an exact expression for $s$ as a function of $r$.
The second relation is a sum rule of the type \equaref{general-sum
rule}.

For the choice of generators $T_e$--$S_1$, we find that these relations are
\begin{align*} s &= \sqrt{3\left(1-\frac{2\eta}{2-r^2}\right)} -1,\nonumber\\
a &= \sqrt{\frac{\eta}{2(1-\eta)}}r \cos\delta,\end{align*}
and we find similar relations for $T_e$--$S_2$
\begin{align*} s &= \sqrt{\frac{6\eta}{2-r^2}} -1,\nonumber\\
a &= -\sqrt{\frac{\eta}{2(1-\eta)}}r \cos\delta.\end{align*}

For $T_\mu$--$S_2$, we see a non-zero prediction for $a$ at $r=0$
\begin{align*} s &= \sqrt{\frac{3(1-\eta)}{2-r^2}} -1,\nonumber\\
a &= \frac{1-3\eta}{2(1+\eta)}-\sqrt{\frac{1-\eta}{2(1+\eta)}}r \cos\delta.
\end{align*}
This feature is also present for the choice  $T_\tau$--$S_2$,
\begin{align*} s &= \sqrt{\frac{3(1-\eta)}{2-r^2}} -1,\nonumber\\
a &= -\frac{1-3\eta}{2(1+\eta)}-\sqrt{\frac{1-\eta}{2(1+\eta)}}r \cos\delta.
\end{align*}
By comparing the predictions of $s$ for all of the choices of $(m,p)$,
$k$ and $d$ with the known phenomenological interval, we identify 8 viable
scenarios which are listed in \tabref{all-sumrules-numbers} together with
their numerical predictions for $s$, $a_0$ and $\lambda$. Analytical
expressions for each of these 8 scenarios can be found in \tabref{tab:all-sumrules}.

\begin{table}[t]
\centering
\begin{tabular}{c c c c c c}
$G_f$ & $m$ & $T_\alpha$,$S_i$ & $s$ & $a_0$ & $\lambda$\\
\hline\hline
\multirow{3}{*}{A$_4$} & $3$ & $T_e$,$S_2$ & $\frac{1}{\sqrt{1-r^2/2}}-1$ & $0$ & $-\frac{1}{2}$ \\
& $3$ & $T_\mu$,$S_2$ & $\frac{1}{\sqrt{1-r^2/2}}-1$ & $0$ & $-\frac{1}{2}$ \\
& $3$ & $T_\tau$,$S_2$ & $\frac{1}{\sqrt{1-r^2/2}}-1$ & $0$ & $-\frac{1}{2}$ \\
\hline
\multirow{3}{*}{S$_4$} & $3$ & $T_e$,$S_1$ & $\sqrt{1-\frac{2r^2}{2-r^2}}-1$ & $0$ & $1$\\
& $4$ & $T_\mu$,$S_2$ & $\sqrt{\frac{3}{2(2-r^2)}}-1$ & $\frac{1}{6}$ & $-\sqrt{\frac{1}{6}}$ \\
& $4$ & $T_\tau$,$S_2$ & $\sqrt{\frac{3}{2(2-r^2)}}-1$ & $-\frac{1}{6}$ & $-\sqrt{\frac{1}{6}}$ \\
\hline
\multirow{4}{*}{A$_5$}
& $5$ & $T_e$,$S_1$ & $\sqrt{3+\frac{6}{\left(3-\varphi\right) \left(r^2-2\right)}}-1$ & $0$ & $\frac{\varphi}{\sqrt{2}}$ \\
& $5$ & $T_e$,$S_2$ & $\sqrt{\frac{6}{(2+\varphi)(2-r^2)}}-1$ & $0$ & $\frac{1-\varphi}{\sqrt{2}}$ \\
& $5$ & $T_\mu$,$S_2$ & $\sqrt{\frac{3 \varphi}{(2\varphi-1)(2-r^2)}}-1$ & $-\frac{5-4\varphi}{22}$ & $-\sqrt{\frac{3+2\varphi}{22}}$ \\
& $5$ & $T_\tau$,$S_2$ & $\sqrt{\frac{3\varphi}{(2\varphi-1)(2-r^2)}}-1$ & $\frac{5-4\varphi}{22}$ & $-\sqrt{\frac{3+2\varphi}{22}}$ \\
\end{tabular}

\caption{\label{tab:all-sumrules}Analytical expressions for the
phenomenologically viable sum rules arising in the Hernandez-Smirnov
framework for finite von Dyck groups, as described in Table I. In this table,
$m$ gives the order of the generator which controls the charge lepton mass
matrix, $T^m_\alpha=1$, and $\varphi=(1+\sqrt{5})/2$ is the golden ratio.}
\end{table}
%

%%%%%%%%%%%%%%%%%%%%%%%%%%%%%%%%
%%%%%%%%%%%%%%%%%%%%%%%%%%%%%%%%
%%%%%%%%%%%%%%%%%%%%%%%%%%%%%%%%

\bibliographystyle{apsrev4-1}
\bibliography{Alexandria}{}
\end{document}